\documentclass[aps,prd,10pt,notitlepage,superscriptaddress,nofootinbib,numbers]{revtex4-2}

\usepackage[utf8]{inputenc}
\usepackage[T1]{fontenc}
\usepackage{lmodern}
\usepackage{microtype}
\usepackage{amsmath,amssymb,amsfonts,mathrsfs}
\usepackage{bm}
\usepackage{graphicx}
\usepackage{float}
\usepackage[dvipsnames]{xcolor}
\usepackage{hyperref}
\hypersetup{colorlinks=true,citecolor=Purple,linkcolor=Purple,urlcolor=Purple,breaklinks=true}
\usepackage[capitalize]{cleveref}
\usepackage{tikz}

\definecolor{lime}{HTML}{A6CE39}
\DeclareRobustCommand{\orcidicon}{%
	\begin{tikzpicture}
		\draw[lime, fill=lime] (0,0) 
		circle [radius=0.16] 
		node[white] {{\fontfamily{qag}\selectfont \tiny ID}};
		\draw[white, fill=white] (-0.0625,0.095) 
		circle [radius=0.007];
	\end{tikzpicture}
	\hspace{-2mm}
}

\foreach \x in {A, ..., Z}{%
	\expandafter\xdef\csname orcid\x\endcsname{\noexpand\href{https://orcid.org/\csname orcidauthor\x\endcsname}{\noexpand\orcidicon}}
}

\newcommand\orcidJonathan{{\href{https://orcid.org/0000-0001-9291-0893}{\orcidicon}}}
\newcommand\orcidEdson{{\href{https://orcid.org/0000-0001-9929-5977}{\orcidicon}}}
\newcommand\orcidPedro{{\href{https://orcid.org/0000-0002-8478-5460}{\orcidicon}}}

\newcommand{\dd}{\mathrm{d}}

\begin{document}

% Presentation only: natural vertical spacing and less restrictive floats.
\raggedbottom
\setcounter{topnumber}{4}
\setcounter{bottomnumber}{3}
\setcounter{totalnumber}{6}
\setlength{\abovecaptionskip}{5pt}
\setlength{\belowcaptionskip}{3pt}
\allowdisplaybreaks[1]
\setlength{\textfloatsep}{8pt plus 2pt minus 2pt}
\setlength{\floatsep}{8pt plus 2pt minus 2pt}
\setlength{\intextsep}{8pt plus 2pt minus 2pt}
\setlength{\abovedisplayskip}{7pt plus 2pt minus 2pt}
\setlength{\belowdisplayskip}{7pt plus 2pt minus 2pt}
\setlength{\abovedisplayshortskip}{4pt plus 1pt minus 1pt}
\setlength{\belowdisplayshortskip}{5pt plus 1pt minus 1pt}
\renewcommand{\topfraction}{0.95}
\renewcommand{\bottomfraction}{0.90}
\renewcommand{\textfraction}{0.05}
\renewcommand{\floatpagefraction}{0.90}
\makeatletter
\setlength{\@fptop}{0pt}
\setlength{\@fpsep}{12pt plus 2pt minus 2pt}
\setlength{\@fpbot}{0pt plus 1fil}
\makeatother

\title{Slowly rotating white dwarfs in \texorpdfstring{$f(R,T)=R+2\lambda T$}{f(R,T)=R+2lambda T} gravity:\texorpdfstring{\\}{ }
structure and stationary tidal response}

\author{Edson Otoniel\orcidEdson\!\!}
\email{edson.otoniel@ufca.edu.br}
\affiliation{Universidade Federal do Cariri (UFCA), Instituto de Forma\c{c}\~ao de Educadores - IFE,  R. Oleg\'ario Emidio de Araujo S/N, Brejo Santo - CE, 63.260-000 - Brazil}

\author{Jonathan A. Rebou\c{c}as\orcidJonathan\!\!}
\email{jalvesreboucas@ifce.edu.br}
\affiliation{Instituto Federal de Educa\c{c}\~ao, Ci\^encia e Tecnologia do Cear\'a (IFCE), Iguatu, CE, Brazil}

\author{P. H. R. S. Moraes\orcidPedro\!\!}
\email{moraes.phrs@gmail.com}
\affiliation{N\'ucleo de Astrof\'isica, Universidade Cidade de S\~ao Paulo, Rua Galv\~ao Bueno 868, S\~ao Paulo 01506-000, Brazil}

\date{\today}

\begin{abstract}
We calculate the structure and stationary tidal response of cold carbon fluid white dwarfs in $f(R,T)=R+2\lambda T$ gravity through second order in uniform rotation. Equilibria are compared at fixed baryon number, with angular momentum also conserved during each aligned electric quadrupolar tidal perturbation. Negative coupling increases radii and dimensional tidal deformabilities over the common mass range. For the baryon sequence studied, background readjustment accounts for 95.8--97.5\% of the rotational enhancement at 40\% of each model's Kepler reference frequency. The explicit mixed rotational tidal contribution is positive and numerically resolved, but subdominant. At that spin, the enhancement relative to each coupling's static model decreases from 107.4\% in general relativity to 14.2\% at the most negative coupling, despite its larger absolute deformability. This reduction reflects less relative background expansion and a different Love-number variation.
\end{abstract}

\maketitle
% Local title-to-contents spacing for the completed abstract.
\vspace{-4pt}
\tableofcontents

\section{Introduction}\label{sec:introduction}

White dwarfs represent the final evolutionary stage of stars of low and intermediate mass, including the Sun. Such compact remnants contain a mass comparable to that of the Sun within a radius similar to Earth’s and are supported against gravitational collapse primarily by electron degeneracy pressure \cite{althaus/2010}. 

These remnants play an important role in cosmology as they are the progenitors of Type Ia supernovae, thermonuclear explosions whose standardized luminosities make them powerful indicators of cosmological distances \cite{maoz/2014}. In explosion scenarios near the Chandrasekhar mass, a white dwarf composed of carbon and oxygen approaches the maximum mass that electron degeneracy pressure can support in a cold, nonrotating configuration, approximately $1.4\,M_\odot$ \cite{chandrasekhar/1931,althaus/2010}. The resulting increase in central density and temperature can trigger thermonuclear runaway \cite{maoz/2014}. Observations of these supernovae revealed that the expansion of the Universe is accelerating, providing foundational evidence for dark energy \cite{riess/1998,perlmutter/1999}.

White dwarfs are expected to play a central role in the future of gravitational wave astronomy, particularly through compact double white dwarf binaries emitting in the millihertz frequency range accessible to the Laser Interferometer Space Antenna (LISA). Population models predict that LISA will individually resolve tens of thousands of these systems, providing data for investigating binary evolution and mapping the Galactic population of compact remnants \cite{korol/2017}. Several already known binaries containing white dwarfs are expected to serve as verification sources, helping to validate the observatory’s performance during its initial operation \cite{kupfer/2024}. Combining gravitational wave measurements with electromagnetic observations will also enable tighter constraints on binary masses, orbital inclinations and distances, allowing the evolution of these systems toward mass transfer or merger to be studied \cite{korol/2017,kupfer/2024}.

These stars also provide laboratories for investigating how the theory of gravity affects stellar structure. Their equilibrium configurations depend on the balance between gravitational attraction and electron degeneracy pressure, allowing modifications to gravity to alter the predicted relation between mass and radius even when the underlying matter equation of state remains unchanged \cite{Carvalho2017}. Related investigations in $f(R)$ gravity have also examined the Chandrasekhar mass limit and the structural properties of white dwarfs \cite{Astashenok2022,Astashenok2023}. These alterations can raise the maximum mass above the conventional Chandrasekhar limit \cite{otoniel/2026,kalita/2021,li/2024,das/2015}, thereby admitting configurations known as super-Chandrasekhar white dwarfs. Such configurations offer a possible explanation for unusually luminous Type Ia supernovae whose inferred progenitor masses exceed this limit \cite{howell/2006,scalzo/2010,taubenberger/2011}.

Rotation and tidal deformation extend this investigation beyond spherical equilibrium: the moment of inertia, rotational quadrupole and tidal deformability characterize complementary aspects of the stellar response to rotational and external gravitational fields \cite{taylor/2020}. Studying these quantities within a consistent gravitational framework can therefore help identify departures from general relativity and assess their degeneracies with the effects of mass, spin and composition.

For the rapidly rotating white dwarf in CTCV J2056--3014, equilibrium calculations connect the spin period to the allowed mass interval~\cite{Otoniel2021Rotation}. Comparisons with rotating strange dwarfs also show that rotation can reduce the separation between different internal compositions in the mass--radius plane~\cite{Otoniel2026StrangeRotation}.

We adopt the $f(R,T)$ theory of gravity \cite{Harko2011} as our underlying scenario. The $f(R,T)$ theory extends the formalism of general relativity by replacing the Ricci scalar $R$ in the gravitational action with a generic function $f$ of both $R$ and the trace $T$ of the stress tensor. Such a modification of the gravitational action yields extra terms in the field equations. Cosmological applications have explored whether such terms can account for the observed accelerated expansion of the Universe \cite{moraes/2016,sahoo/2018,myrzakulov/2025}. 

Compact astrophysical objects, such as neutron stars and strange stars, have also been studied in such a modified gravity formalism \cite{sharif/2018,chowdhury/2025,carvalho/2020,deb/2019,dos_santos/2019}. In particular, white dwarfs have been investigated in $f(R,T)$ gravity \cite{Carvalho2017}. Charged white dwarfs have been studied in Refs.~\cite{feng/2025,rocha/2020}, and their cooling has been investigated in Ref.~\cite{bhattacharjee/2024}. White dwarf equilibrium has also been studied with quadratic teleparallel torsion, a geometrical modification distinct from coupling to the matter trace~\cite{Otoniel2026Torsion}.

Rotation in this framework has already been studied by Murshid and Kalam \cite{MurshidKalam2024}, who used the Hartle--Thorne formalism \cite{Hartle1967,HartleThorne1968} to calculate isolated neutron stars through second order in angular velocity. They also evaluated tidal deformability, assuming a nonspinning star in that calculation.

The effects of rotation on tides have also been investigated. Pani et al.~\cite{Pani2015} obtained the stationary exterior geometry of a spinning object in an axisymmetric tidal field through second order in spin. Fuller and Lai \cite{FullerLai2014} studied the influence of rotation on dynamical tides in white dwarf binaries. These are distinct physical settings: the latter involves fluid motions varying in time, whereas the response considered here is stationary and axisymmetric. Static calculations for crystallized white dwarfs also show that elasticity affects tidal deformability \cite{PerotChamel2022}.

These studies motivate extending the investigation of white dwarfs in $f(R,T)$ gravity beyond spherical equilibrium to include both rotation and tidal deformation. Here, we adopt the model $f(R,T)=R+2\lambda T$, where $\lambda$ is a free parameter, and calculate uniformly rotating configurations through second order in angular velocity, together with their stationary response to an aligned electric quadrupolar tidal field. We compare selected isolated equilibria at fixed physical baryon number and conserve both baryon number and angular momentum during each tidal perturbation. This prescription allows us to examine how the gravitational coupling changes the stellar structure and tidal response while keeping the amount of matter fixed.

A central objective is to distinguish the change in tidal deformability caused by rotational readjustment of the stellar background from the explicit mixed rotational tidal contribution within the adopted perturbative prescription. This distinction is important because a larger deformability need not imply a stronger fractional sensitivity to rotation: changes in the stellar radius and internal structure also affect the response. By examining these contributions separately, we assess how the modified gravitational equilibrium influences the rotational enhancement of tidal deformability. The resulting calculations provide a basis for future dynamical and binary studies aimed at determining whether tidal measurements can distinguish gravitational modifications from variations in stellar properties.

The paper is organized as follows. We present the gravity model, matter description and perturbative formalism in Section~\ref{sec:theoreticalframework}, and discuss the structural and tidal results in Section~\ref{sec:results}. The conclusions and possible extensions are presented in Section~\ref{sec:conclusion}.

\section{Theoretical Framework}\label{sec:theoreticalframework}

\subsection{Gravity model and effective-fluid representation}\label{sec:gravity}

We use the metric signature $(-,+,+,+)$ and geometrized units $G=c=1$
unless dimensional constants are displayed explicitly. Greek indices run
over spacetime coordinates, $g=\det g_{\mu\nu}$, and $\nabla_\mu$ is the
metric covariant derivative. Energy density and pressure then have units
of inverse length squared, mass has units of length, and angular velocity
has units of inverse length. The action of the model is
\begin{equation}
 S_{\rm act}=\frac{1}{16\pi}\int \sqrt{-g}\,(R+2\lambda T)\,\dd^4x
   +\int\sqrt{-g}\,\mathcal L_m\,\dd^4x,
 \label{eq:th-action}
\end{equation}
where $R$ is the Ricci scalar, $T=g^{\mu\nu}T_{\mu\nu}$ is the trace of the
physical matter stress tensor, and $\lambda$ is a dimensionless constant.
This is the linear trace-coupling specialization of $f(R,T)$ gravity
\cite{Harko2011}. We describe matter as an isotropic fluid,
\begin{equation}
 T_{\mu\nu}=(\epsilon+p)u_\mu u_\nu+p g_{\mu\nu},\qquad 
u^\mu u_\mu=-1,\qquad T=-\epsilon+3p,
\label{eq:th-physical-fluid}
\end{equation}
with physical energy density $\epsilon$, pressure $p$, and four-velocity
$u^\mu$. The matter prescription adopted here is $\mathcal L_m=p$, with
$\Theta_{\mu\nu}\equiv g^{\alpha\beta}\delta T_{\alpha\beta}/\delta
 g^{\mu\nu}=-2T_{\mu\nu}+p g_{\mu\nu}$. Specifying this prescription is
necessary: in a theory with explicit $T$ dependence, alternative choices
for the metric variation of the matter sector need not give the same
stellar equations. Related calculations of white dwarf equilibrium in
$f(R,T,\mathcal L_m)$ gravity explicitly compare different matter
Lagrangians~\cite{otoniel/2026}.

To obtain the field equations, the trace varies according to
$\delta T=(T_{\mu\nu}+\Theta_{\mu\nu})\delta g^{\mu\nu}$. After integrating
the curvature variation by parts, the general metric equation is
\begin{equation}
 f_R R_{\mu\nu}-\frac12 f g_{\mu\nu}
 +(g_{\mu\nu}\Box-\nabla_\mu\nabla_\nu)f_R
 =8\pi T_{\mu\nu}-f_T(T_{\mu\nu}+\Theta_{\mu\nu}),
 \label{eq:th-general-field}
\end{equation}
where $R_{\mu\nu}$ is the Ricci tensor, $f_R=\partial f/\partial R$,
$f_T=\partial f/\partial T$, and $\Box=\nabla_\alpha\nabla^\alpha$.
Inserting $f_R=1$, $f_T=2\lambda$, and
the specified $\Theta_{\mu\nu}$ gives
\begin{equation}
 G_{\mu\nu}=(8\pi+2\lambda)T_{\mu\nu}
             +\lambda(T-2p)g_{\mu\nu}
           \equiv8\pi T^{\rm e}_{\mu\nu}.
 \label{eq:th-field}
\end{equation}
Here $G_{\mu\nu}=R_{\mu\nu}-Rg_{\mu\nu}/2$ is the Einstein tensor.
The superscript e denotes an effective gravitational source, not an
additional material component. Related effective-fluid
descriptions of modified gravity have been explored in earlier
cosmological studies \cite{NojiriOdintsov2006,Capozziello2006}.
Define
\begin{equation}
 a=\frac{\lambda}{8\pi},\qquad b=1+3a,\qquad
 \epsilon_{\rm e}=b\epsilon-a p,\qquad
 p_{\rm e}=b p-a\epsilon.
\label{eq:th-map}
\end{equation}
Collecting the coefficients of $u_\mu u_\nu$ and $g_{\mu\nu}$ in
Eq.~\eqref{eq:th-field} yields
\begin{equation}
 T^{\rm e}_{\mu\nu}=(\epsilon_{\rm e}+p_{\rm e})u_\mu u_\nu
                     +p_{\rm e}g_{\mu\nu},\qquad 
\epsilon_{\rm e}+p_{\rm e}=(1+2a)(\epsilon+p).
\label{eq:th-effective-fluid}
\end{equation}
This algebraic representation applies at every retained perturbative
order, provided the same matter prescription is used. The Bianchi identity
implies $\nabla^\mu T^{\rm e}_{\mu\nu}=0$, whereas the physical tensor obeys
\begin{equation}
 \nabla^\mu T_{\mu\nu}
 =\frac{\lambda}{8\pi+2\lambda}\nabla_\nu(\epsilon-p).
 \label{eq:th-nonconservation}
\end{equation}
Thus the pressure gradient differs from that of a GR star with the same
physical EOS. In vacuum both matter variables vanish and
Eq.~\eqref{eq:th-field} reduces to the vacuum Einstein equation; there is
no extra propagating curvature scalar associated with $f_R$ in this model.
The spherical white-dwarf application of this coupling was studied in
Ref.~\cite{Carvalho2017}.

The effective barotrope must be differentiated consistently. Writing
$\chi=\dd\epsilon/\dd p$ and
$D=\dd p_{\rm e}/\dd p=b-a\chi$, the chain rule gives
\begin{subequations}\label{eq:th-eos-derivatives}
\begin{align}
 \chi_{\rm e}\equiv\frac{\dd\epsilon_{\rm e}}{\dd p_{\rm e}}
  &=\frac{b\chi-a}{D},\\
 \frac{\dd^2\epsilon_{\rm e}}{\dd p_{\rm e}^2}
  &=\frac{(b^2-a^2)\,\dd^2\epsilon/\dd p^2}{D^3},
\end{align}
\end{subequations}
\begin{subequations}\label{eq:th-baryon-derivatives}
\begin{align}
 \frac{\dd n_B}{\dd p_{\rm e}}&=\frac{\dd n_B/\dd p}{D},\\
 \frac{\dd^2 n_B}{\dd p_{\rm e}^2}
  &=\frac{D\,\dd^2 n_B/\dd p^2
       +a(\dd n_B/\dd p)(\dd^2\epsilon/\dd p^2)}{D^3}.
\end{align}
\end{subequations}
Here $n_B$ is the physical baryon number density. We require $D>0$ so
that effective pressure labels the physical EOS uniquely. These expressions
are local identities for a differentiable barotrope. For a tabulated EOS,
the derivatives entering the perturbation equations are those of the
actual interpolated effective barotrope, with one-sided values at its
knots. No effective baryon density is introduced.

\subsection{Cold carbon matter and spherical equilibrium}\label{sec:matter-static}
\subsubsection{Electron degeneracy, ionic matter, and lattice contribution}

We keep the composition fixed to $^{12}\mathrm C$ and describe the
material by an isotropic EOS at zero temperature and zero magnetic field.
If $n_i$, $n_e$, and $n_B$ denote the ion, electron, and baryon number
densities, charge neutrality fixes $n_e=Zn_i=(Z/A_{\rm ion})n_B$,
with $Z=6$ and $A_{\rm ion}=12$. Thus there is one electron for every two
baryons. The nuclear mass number $A_{\rm ion}$ is distinct from the total
stellar baryon number $A$. As in the gravitational equations, we use
$G=c=1$, while retaining the reduced Planck constant $\hbar$ explicitly.
The local electron distribution fills momentum states up to the Fermi
momentum $p_F$, giving
\begin{subequations}\label{eq:th-fermi}
\begin{align}
 n_e&=\frac{p_F^3}{3\pi^2\hbar^3},\\
E(q)&=\sqrt{m_e^2+q^2},\\
\epsilon_{\rm el}&=\frac{1}{\pi^2\hbar^3}
                       \int_0^{p_F} E(q)q^2\,\dd q,\\
p_{\rm el}&=\frac{1}{3\pi^2\hbar^3}
                       \int_0^{p_F}\frac{q^4}{E(q)}\,\dd q.
\end{align}
\end{subequations}
Here $q$ is the integration momentum and $m_e$ is the electron rest mass;
the chemical potential $\mu_e=E(p_F)$ includes rest energy. Electrons
supply the degeneracy pressure, whereas the ionic rest energy dominates
the material energy density over the white dwarf regime. For
$p_F\ll m_e$ the electron pressure scales as $n_e^{5/3}$; for
$p_F\gg m_e$ it approaches an $n_e^{4/3}$ dependence. The full integrals
cover the transition between these limits without imposing either
polytropic approximation \cite{Chandrasekhar1931}.

Coulomb interactions reduce both the energy and the pressure at a given
composition and number density. In Gaussian electromagnetic conventions,
the leading lattice energy per ion scales as $-Z^2e^2/a_i$, where $e$ is
the elementary charge and $a_i=(3/4\pi n_i)^{1/3}$ is the ion sphere
radius. A density independent lattice coefficient therefore gives
$\epsilon_L\propto-n_i^{4/3}$ and
\begin{equation}
 p_L=n_i\frac{\dd\epsilon_L}{\dd n_i}-\epsilon_L
     =\frac{\epsilon_L}{3}.
 \label{eq:th-lattice}
\end{equation}
This is the electrostatic correction to the degenerate electron gas
\cite{Salpeter1961}. The total physical pressure $p$ and energy density
$\epsilon$, which enter the right-hand side of Eq.~\eqref{eq:th-map}, are
\begin{equation}
 p=p_{\rm el}+p_L,\qquad 
 \epsilon=\epsilon_{\rm ion}+\epsilon_{\rm el}+\epsilon_L,
\label{eq:th-carbon-total}
\end{equation}
where $\epsilon_{\rm ion}$ contains the ionic rest energy. For pressure
and energy density, the subscripts el and e distinguish the electron
contribution from the effective source obtained by applying
Eq.~\eqref{eq:th-map} to these material totals. Since
$\epsilon_{\rm el}$ already includes electron rest energy, a rest energy
constructed from neutral atomic masses must have that electronic
contribution removed before it enters $\epsilon_{\rm ion}$.
At fixed composition and zero temperature, these contributions define
the material barotrope $p(\epsilon)$, with mass density $\rho=\epsilon$
in the chosen units. We treat the matter as an isotropic
fluid: the Coulomb lattice contributes to its pressure and energy
density, while elastic shear stresses are neglected.

Compression eventually makes a fixed carbon composition untenable.
Electron capture changes the nuclear charge and hence the number of
pressure supporting electrons per baryon. For a single capture,
$(A_{\rm ion},Z)+e^-\rightarrow(A_{\rm ion},Z-1)+\nu_e$,
the threshold without medium corrections is
\begin{equation}
 \mu_e^{\beta,0}=m_e+M_{\rm at}(A_{\rm ion},Z-1)
                         -M_{\rm at}(A_{\rm ion},Z),
 \label{eq:th-capture-threshold}
\end{equation}
where $M_{\rm at}$ denotes a neutral atomic mass and differences in
atomic electron binding energies are neglected. Neutrinos are assumed
to escape. In the medium, the transition is determined by comparing the
parent and daughter Gibbs energies per baryon,
$\mu_B=(\epsilon+p)/n_B$, at the same physical pressure, including their
lattice energies \cite{ChamelFantina2015Capture}. This criterion concerns
the physical EOS, not the effective source pressure $p_{\rm e}$.
For carbon without a magnetic field, the reference capture threshold is
$\rho_\beta\simeq4.16\times10^{10}\ \mathrm{g\,cm^{-3}}$
\cite{ChamelFantina2015Capture}. This threshold limits the regime in
which the assumed carbon composition can be maintained.

Neutron drip is a separate change of composition, characterized by the
appearance of unbound neutrons. Its location depends on nuclear masses
and the conditions of matter equilibrium \cite{Chamel2015NeutronDrip}.
The present description is restricted to carbon nuclei and electrons,
with all neutrons bound within nuclei. In this regime, the coupling
$\lambda$ changes gravitational equilibrium through
Eq.~\eqref{eq:th-map}, while the physical EOS remains fixed.

\subsubsection{Static structure equations and central regularity}

For a spherical configuration let $r$ be the areal radius and write
\begin{equation}
 \dd s_0^2=-e^{\nu(r)}\dd t^2+\frac{\dd r^2}{w(r)}
                 +r^2\dd\Omega_2^2,\qquad 
w=1-\frac{2m(r)}{r},\qquad 
\dd\Omega_2^2=\dd\theta^2+\sin^2\theta\,\dd\phi^2.
\label{eq:th-static-metric}
\end{equation}
For the enclosed gravitational mass $m(r)$, the $t$--$t$ field equation
gives the first structure equation. The radial equation determines $\nu'$, and the radial
component of effective stress conservation determines $p_{\rm e}'$:
\begin{subequations}\label{eq:th-tov}
\begin{align}
 m'&=4\pi r^2\epsilon_{\rm e},\\
\frac{\nu'}{2}\equiv g_0&=\frac{m+4\pi r^3p_{\rm e}}{r^2w},\\
p_{\rm e}'&=-(\epsilon_{\rm e}+p_{\rm e})g_0,\\
p'&=-\frac{(1+2a)(\epsilon+p)}{D}\,
             \frac{m+4\pi r^3p_{\rm e}}{r^2w}.
\end{align}
\end{subequations}
A prime now denotes $\dd/\dd r$ unless another derivative is written
explicitly. In particular, $\chi_{\rm e}$ always denotes a pressure
derivative, not a radial derivative. The quantity $g_0$ has units of
inverse length and measures the radial gravitational potential gradient.
Setting $\lambda=0$ recovers the GR hydrostatic equations.

Central regularity fixes $m(0)=0$ and leaves a central physical pressure
$p_c$ as the equilibrium parameter. Denoting central effective values by
$\epsilon_{{\rm e}c}$ and $p_{{\rm e}c}$, expansion of Eq.~\eqref{eq:th-tov}
gives
\begin{equation}
\begin{aligned}
 m(r)&=\frac{4\pi}{3}\epsilon_{{\rm e}c}r^3+O(r^5),\\
 p_{\rm e}(r)&=p_{{\rm e}c}
 -2\pi(\epsilon_{{\rm e}c}+p_{{\rm e}c})
             (\epsilon_{{\rm e}c}/3+p_{{\rm e}c})r^2+O(r^4),\\
 \nu(r)&=\nu_c+4\pi(\epsilon_{{\rm e}c}/3+p_{{\rm e}c})r^2+O(r^4).
\end{aligned}
\label{eq:th-center}
\end{equation}
Exterior time normalization fixes the additive constant $\nu_c$.
The spherical gravitational mass and radius are denoted by $M_0$ and
$R_0$, reserving $M$ for a rotating equilibrium.

\subsubsection{Operational surface and vacuum matching}

Integration terminates at the declared low-pressure endpoint of the
EOS, defining $R_0$ and $M_0=m(R_0)$. At an exact fluid--vacuum boundary
without a surface stress tensor, the normal effective pressure must
vanish. The present EOS cutoff leaves a small residual $p_{\rm e}(R_0)$;
its contribution is neglected in the vacuum boundary prescription, with
its size monitored through
\begin{equation}
 q_s=\frac{4\pi R_0^3p_{\rm e}(R_0)}{M_0}.
 \label{eq:th-surface-pressure}
\end{equation}
This dimensionless ratio compares the pressure term and mass term in
the surface value of $g_0$. The lapse is normalized by
$e^{\nu(R_0)}=1-2M_0/R_0$. No atmosphere, envelope, or material thin shell
is added. This prescription does not assert an exact finite-pressure
junction to vacuum. A finite surface density, which can remain at the
cutoff even when pressure is negligible, is retained in the relevant
surface contributions. Pressure and density corrections play different
roles and cannot be interchanged. The corresponding displacement terms
for rotation and tides are specified below.

\subsection{Uniform rotation through second order}\label{sec:rotation}
\subsubsection{First-order frame dragging and angular momentum}

We adopt uniform circular flow,
$u^\mu=u^t(1,0,0,\Omega)$, with constant stellar angular velocity
$\Omega$. In the Hartle gauge the circular metric is
\begin{equation}
 \dd s^2={}-e^\nu(1+2h)\dd t^2
 +\frac{1+2\mathfrak m/(rw)}{w}\dd r^2 
 +r^2(1+2k)\{\dd\theta^2+
                    \sin^2\theta[\dd\phi-\omega\dd t]^2\}.
\label{eq:th-circular-metric}
\end{equation}
For an isolated rotating star, $\omega=O(\Omega)$ is the dragging angular
velocity, while
$h=h_0+h_2P_2(\cos\theta)$,
$\mathfrak m=m_0+m_2P_2(\cos\theta)$, and
$k=k_2^{\rm met}P_2(\cos\theta)$ are $O(\Omega^2)$.
The $P_\ell$ are Legendre polynomials. The radial gauge sets the monopole
of $k$ to zero. The metric function $k_2^{\rm met}$ is distinct from the
tidal Love number $k_2$. Expanding Eq.~\eqref{eq:th-field} in this metric
is equivalent to the Hartle equations with effective fluid variables
\cite{Hartle1967,HartleThorne1968}.

Define the fluid angular velocity relative to local inertial frames by
$\bar\omega=\Omega-\omega$ and
$j=e^{-\nu/2}\sqrt{w}$. The first-order $t$--$\phi$ equation becomes
\begin{equation}
 (r^4j\bar\omega')'+4r^3j'\bar\omega=0.
 \label{eq:th-frame}
\end{equation}
Its regular solution has finite $\bar\omega(0)$ and
$\bar\omega'(0)=0$. The normalization is fixed by the prescribed
$\Omega$. Since the vacuum solution is
$\bar\omega=\Omega-2J/r^3$, matching at $R_0$ gives
\begin{subequations}\label{eq:th-spin}
\begin{align}
 J&=\frac{R_0^4}{6}\bar\omega'(R_0),\\
\Omega&=\bar\omega(R_0)+\frac{2J}{R_0^3},\\
I&=\frac{J}{\Omega}.
\end{align}
\end{subequations}
Here $J$ is the total angular momentum, with units of length squared,
and $I$ is the leading moment of inertia, with units of length cubed.
First-order dragging is needed in the second-order source terms even
when only the mass or surface shape is being considered.

\subsubsection{Mass correction, shape, and rotational quadrupole}

Introduce the enthalpy perturbations
$p_\ell^*=\delta p_{{\rm e},\ell}/(\epsilon_{\rm e}+p_{\rm e})$.
For a rotating model paired with a spherical star at the same central
pressure, $p_0^*(0)=0$. The monopole field equations are
\begin{subequations}\label{eq:th-rotation-monopole}
\begin{align}
 m_0'={}&4\pi r^2\chi_{\rm e}(\epsilon_{\rm e}+p_{\rm e})p_0^*
 +\frac{r^4j^2}{12}(\bar\omega')^2
 -\frac{r^3}{3}(j^2)'\bar\omega^2,\\
(p_0^*)'={}&-\frac{1+8\pi r^2p_{\rm e}}{r^2w^2}m_0
 -\frac{4\pi r(\epsilon_{\rm e}+p_{\rm e})}{w}p_0^* +\frac{r^3j^2}{12w}(\bar\omega')^2
 +\frac13\left(\frac{r^2j^2\bar\omega^2}{w}\right)'.
\end{align}
\end{subequations}
In $m_0'$, the material term represents the change of the effective energy density;
the remaining terms are sourced by rotation and dragging. The same
Euler equation gives
\begin{subequations}\label{eq:th-rotation-euler}
\begin{align}
 h_0&=-p_0^*+\frac{r^2e^{-\nu}\bar\omega^2}{3}+C_s,\\
p_2^*&=-h_2-\frac{r^2e^{-\nu}\bar\omega^2}{3},
\end{align}
\end{subequations}
where $C_s$ is an integration constant determined by matching.

For the quadrupole system, use $v_2=h_2+k_2^{\rm met}$. Define radial coefficients
\begin{equation}
 \mathcal A=-\nu'+\frac{8\pi(\epsilon_{\rm e}+p_{\rm e})-4m/r^3}{w\nu'},\qquad 
\mathcal B_\ell=-\frac{\ell(\ell+1)-2}{r^2w\nu'}.
\label{eq:th-polar-coefficients}
\end{equation}
Combining the angular and radial constraints gives
\begin{subequations}\label{eq:th-rotation-quadrupole}
\begin{align}
 h_2'&=\mathcal A h_2+\mathcal B_2v_2+S_h,\\
v_2'&=-\nu'h_2+S_v,\\
S_h&=\frac{r^3j^2(\bar\omega')^2}{6}
          \left(\frac{r\nu'}2-\frac{1}{rw\nu'}\right) -\frac{r^2(j^2)'\bar\omega^2}{3}
          \left(\frac{r\nu'}2+\frac{1}{rw\nu'}\right),\\
S_v&=\left(\frac1r+\frac{\nu'}2\right)
       \left[\frac{r^4j^2(\bar\omega')^2}{6}
                   -\frac{r^3(j^2)'\bar\omega^2}{3}\right],\\
m_2&=rw\left[-h_2-\frac{r^3(j^2)'\bar\omega^2}{3}
                       +\frac{r^4j^2(\bar\omega')^2}{6}\right].
\end{align}
\end{subequations}
A regular particular solution and a regular homogeneous solution are
combined so that the interior matches the asymptotically flat exterior.
These equations generalize to the present matter source without changing
the vacuum metric. A related second-order treatment of neutron stars in
this gravity model is given in Ref.~\cite{MurshidKalam2024}.

An isobar originally at $r$ is displaced by
$\xi(r,\theta)=\xi_0(r)+\xi_2(r)P_2(\cos\theta)$.
The condition that its Lagrangian pressure change vanishes gives
\begin{equation}
 \xi_\ell=-\frac{(\epsilon_{\rm e}+p_{\rm e})p_\ell^*}{p_{\rm e}'}
          =\frac{p_\ell^*}{g_0}.
 \label{eq:th-displacement}
\end{equation}
Evaluating the intrinsic angular scale of the displaced surface gives
the radius convention used for the shape plots,
\begin{subequations}\label{eq:th-radii}
\begin{align}
 R_{\rm eq}&=R_0+\xi_0(R_0)
               -\tfrac12[\xi_2(R_0)+R_0k_2^{\rm met}(R_0)],\\
R_{\rm p}&=R_0+\xi_0(R_0)
                    +\xi_2(R_0)+R_0k_2^{\rm met}(R_0),\\
\mathcal D&=\frac{R_{\rm eq}-R_{\rm p}}{R_0}.
\end{align}
\end{subequations}
These surface radii include the angular metric correction. In particular,
$R_{\rm p}$ is not defined as an integral of proper radial distance.

Writing $M=M_0+\delta M$, the exterior monopole matching condition,
including the displacement of a finite surface density, is
\begin{equation}
 \delta M=m_0(R_0)+\frac{J^2}{R_0^3}
                   +4\pi R_0^2\epsilon_{\rm e}(R_0)\xi_0(R_0).
 \label{eq:th-mass-correction}
\end{equation}
The last term is required when the surface energy density does not
vanish \cite{ReinaVera2015}. It vanishes for an EOS reaching zero density
at its boundary. The isolated structural plots use the original reduced
expression $\delta M=m_0(R_0)+J^2/R_0^3$, whereas the coherent background
of the tidal extension retains the surface term in
Eq.~\eqref{eq:th-mass-correction}; both calculations retain their stated
surface conventions. The rotational quadrupole $Q$ is obtained from
the isolated exterior, with the positive oblateness convention used in
the structure figures. The invariant mass moment in this convention is
$M_2^{\rm rot}=-Q$. It is present at zero applied tide and is separate
from the induced tidal moment.

\subsubsection{Kepler reference and perturbative domain}

At the equator, the circular geodesic equation provides the reference
angular velocity for mass shedding. For a circular orbit with angular velocity
$\Omega_{\rm orb}$, the radial geodesic equation is
$\partial_r g_{tt}+2\Omega_{\rm orb}\partial_r g_{t\phi}
 +\Omega_{\rm orb}^2\partial_r g_{\phi\phi}=0$. Its prograde solution is
\begin{equation}
 \Omega_{\rm orb}=
 \frac{-\partial_r g_{t\phi}+
 \sqrt{(\partial_r g_{t\phi})^2
             -\partial_r g_{tt}\,\partial_r g_{\phi\phi}}}
 {\partial_r g_{\phi\phi}}.
 \label{eq:th-kepler}
\end{equation}
Imposing $\Omega=\Omega_{\rm orb}$ at the rotating equatorial surface
determines the Kepler reference $\Omega_K$ within the second-order metric. We use $x=\Omega/\Omega_K$, with each model's own
reference, and characterize rotational support by
\begin{equation}
 \varepsilon_{\rm rot}=\frac{\Omega^2R_0^3}{M_0}.
 \label{eq:th-spin-domain}
\end{equation}
For a spherical Newtonian star $\Omega_K^2\simeq M_0/R_0^3$, but the
actual reference includes the calculated exterior and shape corrections.
The expansion is controlled by small rotational corrections; using the
truncated metric up to $x=1$ does not supply the omitted $O(\Omega^4)$
even-parity terms. The high-spin structure curves and the restricted
rotation domain used for tidal results must therefore be distinguished.
The rotation period is $P=2\pi/\Omega$ in the same time units as $\Omega$.

\subsection{Physical baryon number and sequence identification}\label{sec:baryons}

We write the physical baryon current as $N^\mu=n_Bu^\mu$. For the stationary,
axisymmetric, circular configurations considered here,
$\nabla_\mu N^\mu=0$: the only nonzero components are in the $t$ and
$\phi$ directions, on which the fields do not depend. This statement
neither replaces Eq.~\eqref{eq:th-nonconservation} nor supplies a
constitutive law for time-dependent perturbations. With
$\mu=\cos\theta$ and a surface $r=R_\Sigma(\mu)$, the invariant count on a
constant-time hypersurface and the conserved rotational charge of the
effective source are
\begin{subequations}\label{eq:th-charges}
\begin{align}
 A&=2\pi\int_{-1}^{1}\dd\mu\int_0^{R_\Sigma(\mu)}
                  n_Bu^t\sqrt{-g}\,\dd r,\\
J&=2\pi\int_{-1}^{1}\dd\mu\int_0^{R_\Sigma(\mu)}
       (\epsilon_{\rm e}+p_{\rm e})u^tu_\phi\sqrt{-g}\,\dd r.
\end{align}
\end{subequations}
Here $g$ is evaluated in coordinates $(t,r,\mu,\phi)$; this accounts for
the angular Jacobian. The baryon number is dimensionless. A baryonic mass
would require an explicitly chosen reference mass per baryon and is not
identified with the gravitational mass. At zero spin,
\begin{equation}
 A_0=4\pi\int_0^{R_0}\frac{n_B(r)r^2}{\sqrt{w(r)}}\,\dd r.
 \label{eq:th-static-baryons}
\end{equation}
In this expression, the density is the tabulated physical $n_B$ composed
with $p_{\rm e}(p)$, not a density inferred by imposing an effective-fluid
first law.

A sequence at prescribed $p_c$ changes its matter content as rotation is
added. To compare isolated equilibria with the same $A$, let $p_{c,0}$
label the initial static model and solve
$A(p_{c,\Omega},\Omega)=A_0(p_{c,0})$. Locally, expanding through second
order gives
\begin{equation}
 p_{c,\Omega}=p_{c,0}+\delta p_c,\qquad 
\delta p_c=-\frac{\delta A_{\rm rot}(p_{c,0},\Omega)}
                         {\dd A_0/\dd p_c}+O(\Omega^4),
\label{eq:th-fixed-baryons}
\end{equation}
provided $\dd A_0/\dd p_c\ne0$. In practice the constraint can instead be
solved directly using the retained baryon integral. The central-density
adjustment identifies a family of equilibria; it is not a calculation
of the time required for spin evolution. Angular momentum varies along
this isolated family. In contrast, when differentiating an equilibrium
with respect to an applied tidal field we impose $\delta A=\delta J=0$.
Fixing $J$ in this derivative generally requires a change in $\Omega$.

\subsection{Rotational corrections to the stationary quadrupolar tidal response}\label{sec:tides}

We calculate the electric quadrupolar response of a uniformly rotating
star, to linear order in the applied tidal field and through second
order in $\Omega$. Stationarity means that the fields have no explicit
time dependence; axial symmetry selects the azimuthal component $m=0$.
The matter rotates with angular velocity $\Omega$, and its response
includes the couplings at $O(\mathcal E\Omega)$ and
$O(\mathcal E\Omega^2)$. During the tidal perturbation the physical
baryon number $A$ and angular momentum $J$ are held fixed. The spherical
problem provides the normalization and the limit of vanishing rotation.

\subsubsection{Spherical reference and tidal normalization}

A tidal field is the spatial variation of an external gravitational
acceleration across the star. In a local description of weak gravity its
quadrupolar potential is $U_{\rm ext}=\mathcal E_{ij}x^ix^j/2$, where
Latin indices label local Cartesian spatial coordinates and
$\mathcal E_{ij}$ is a symmetric electric tidal tensor with zero trace. We
choose the spin as the $z$ axis and write
$\mathcal E_{ij}=\mathrm{diag}(-\mathcal E/2,-\mathcal E/2,\mathcal E)$.
Thus $U_{\rm ext}=\mathcal E r^2P_2(\mu)/2$ and the leading applied term
in $g_{tt}$ is $-\mathcal E r^2P_2(\mu)$. In geometrized units
$[\mathcal E]=L^{-2}$. Linear response is the derivative at vanishing
field, with $|\mathcal E|R_0^3/M_0\ll1$ for a finite imposed deformation.
For a spherical star the induced mass quadrupole with zero trace
satisfies
\begin{subequations}\label{eq:th-love-definitions}
\begin{align}
 Q_{ij}^{\rm ind}&=-\lambda_{\rm T}\mathcal E_{ij},\\
\lambda_{\rm T}&=\frac23 k_2R_0^5,\\
\Lambda_{\rm T}&=\frac{\lambda_{\rm T}}{M_0^5}.
\end{align}
\end{subequations}
The dimensionless electric Love number $k_2$, dimensional deformability
$\lambda_{\rm T}$, and dimensionless deformability $\Lambda_{\rm T}$
are different quantities \cite{Hinderer2008,BinningtonPoisson2009}.
Restoring SI units gives $\lambda_{\rm T}^{\rm SI}=2k_2R_0^5/(3G)$
when $\mathcal E_{ij}$ is the Hessian of the dimensional Newtonian
potential. The geometrized coefficient used in the figures is
$\lambda_{\rm T}=G\lambda_{\rm T}^{\rm SI}$, in units of $\mathrm{km}^5$.

Equilibrium deformation can be distinguished from tidal oscillations
in the Newtonian limit of GR. This comparison defines
the role of the stationary coefficient; it does not introduce a
dynamical calculation in $f(R,T)$ gravity. Here $t$ is dimensional time,
$\rho$ is mass density,
and $\mathcal E_{ij}=\partial_i\partial_j\Phi_{\rm ext}$ has units of
inverse time squared, with $\Phi_{\rm ext}$ the dimensional external
potential. For a small Lagrangian displacement $\boldsymbol\xi$ of a
nonrotating fluid, mass conservation, adiabatic compression, and the
perturbation of its own gravitational field give
\begin{subequations}\label{eq:th-forced-fluid}
\begin{align}
 \delta\rho&=-\boldsymbol\nabla\!\cdot(\rho\boldsymbol\xi),\\
\delta p&=-\boldsymbol\xi\!\cdot\boldsymbol\nabla p
                       -\Gamma_1p\boldsymbol\nabla\!\cdot\boldsymbol\xi,\\
\nabla^2\delta\Phi&=4\pi G\delta\rho,\\
\rho\ddot{\boldsymbol\xi}+\mathcal C_0[\boldsymbol\xi]
   &=-\rho\boldsymbol\nabla U_{\rm ext}.
\end{align}
\end{subequations}
Here $\delta$ denotes an Eulerian perturbation, $\Phi$ is the unperturbed
Newtonian potential, and
$\Gamma_1=(\partial\ln p/\partial\ln\rho)_{s_{\rm ent},X_i}$ holds entropy
$s_{\rm ent}$ and composition fractions $X_i$ fixed.
The restoring operator is
$\mathcal C_0[\boldsymbol\xi]=\boldsymbol\nabla\delta p
 +\delta\rho\boldsymbol\nabla\Phi+\rho\boldsymbol\nabla\delta\Phi$.
The solution is regular at the center, has vanishing Lagrangian
pressure change at a free surface, and matches the exterior perturbed
potential.

For a stable adiabatic nonrotating reference star, the eigenvalue problem
$\mathcal C_0[\boldsymbol\xi_\alpha]
 =\rho\omega_\alpha^2\boldsymbol\xi_\alpha$ defines mode displacements
$\boldsymbol\xi_\alpha$, angular eigenfrequencies $\omega_\alpha$, and
inertias $I_\alpha=\int\rho|\boldsymbol\xi_\alpha|^2\dd^3x$. Expand
$\boldsymbol\xi=\sum_\alpha q_\alpha\boldsymbol\xi_\alpha$ in an
orthogonal discrete mode basis, choosing
$\boldsymbol\xi_\alpha$ to have units of length and $q_\alpha$ to be
dimensionless, and define the
quadrupole overlap
$\mathcal Q_\alpha=\int\rho\boldsymbol\xi_\alpha^*\!\cdot
 \boldsymbol\nabla(r^2P_2)\dd^3x$. Projection then gives
\begin{equation}
 \ddot q_\alpha+2\gamma_\alpha\dot q_\alpha+\omega_\alpha^2q_\alpha
       =-\frac{\mathcal E(t)\mathcal Q_\alpha}{2I_\alpha}.
 \label{eq:th-mode-forcing}
\end{equation}
Weak diagonal damping is parametrized by $\gamma_\alpha$ for this
comparison. Mode expansions for rotating fluids are developed in
Ref.~\cite{Schenk2002}.
The quantities $\omega_\alpha$, $\gamma_\alpha$, and the forcing angular
frequency $\sigma$ have units of inverse time. For forcing proportional
to $e^{-i\sigma t}$, integration by parts in
the induced moment gives
$\delta Q_{zz}=(2/3)\sum_\alpha q_\alpha\mathcal Q_\alpha^*$, and hence
\begin{equation}
 \lambda_{20}^{\rm N}(\sigma)=\frac13\sum_\alpha
 \frac{|\mathcal Q_\alpha|^2}
 {I_\alpha(\omega_\alpha^2-\sigma^2-2i\gamma_\alpha\sigma)}.
 \label{eq:th-mode-response}
\end{equation}
The superscript N identifies the Newtonian expression, with
$[\lambda_{20}^{\rm N}]$ equal to mass times length squared times time
squared. The sum includes modes coupled to the chosen electric
quadrupole, and $1/3$ follows from the tensor normalization. Integration
by parts includes any contribution from the surface density.
When the limit is regular, $\sigma\to0$ gives the equilibrium
compliance. The response near $|\sigma|=\omega_\alpha$ also depends on
the mode frequencies and damping, which a static Love number alone
does not determine.

In a rotating frame the equation for harmonic forcing contains the
Coriolis term $-2i\sigma\rho\boldsymbol\Omega\times\boldsymbol\xi$
and a restoring operator modified by centrifugal distortion. For an
aligned circular binary the forcing frequency in the rotating frame is
$\sigma_m=m(\Omega_{\rm orb}-\Omega)$, where $m$ is the azimuthal harmonic
index. Rotation affects both the modes and the equilibrium structure \cite{FullerLai2014}. Buoyancy modes require
stratification: in a Newtonian spherical fluid their local restoring
frequency obeys
\begin{equation}
 N^2=g_{\rm N}\left(\frac{1}{\Gamma_1}\frac{\dd\ln p}{\dd r}
                         -\frac{\dd\ln\rho}{\dd r}\right),\qquad 
g_{\rm N}=\frac{Gm(r)}{r^2}.
\label{eq:th-buoyancy}
\end{equation}
A barotropic fluid without stratification has $N^2=0$ when $\Gamma_1$
equals the equilibrium adiabatic derivative. Thermal and composition
gradients in white dwarfs can support gravity waves
\cite{FullerLai2012}. Pressure and gravitational forces support acoustic
and fundamental modes, with characteristic scale
$\omega_{\rm dyn}=(GM_0/R_0^3)^{1/2}$; this scale is not a calculated
eigenfrequency. Inertial modes depend on rotation, while nonlinear
couplings can produce tidal dissipation \cite{YuWeinbergFuller2020}.

For the chosen aligned electric component $\ell=2,m=0$, the frequency
$\sigma_m$ vanishes in a circular binary, although the field amplitude
can change on the secular inspiral timescale. The components $m=\pm2$
generally vary with time. Equations~\eqref{eq:th-mode-forcing} and
\eqref{eq:th-mode-response} place the stationary response in this
context; no mode frequencies, damping rates, or dynamical binary signal
are calculated for the present EOS and gravity model.

Returning to geometrized units for the relativistic static calculation,
write the quadrupolar
lapse perturbation as $\delta g_{tt}=-e^\nu H(r)P_2(\mu)$.
An overall sign or normalization of $H$ drops out of its logarithmic
derivative. The static Euler equation gives
$\delta p_{\rm e}=-(\epsilon_{\rm e}+p_{\rm e})H P_2/2$.
Eliminating the other polar metric functions from the linearized Einstein
equations gives $H''+(F+1)H'/r+\mathcal V H=0$; equivalently, with
$y=rH'/H$,
\begin{subequations}\label{eq:th-tidal-ode}
\begin{align}
 ry'+y^2+Fy+r^2\mathcal V&=0,\\
F&=\frac{1+4\pi r^2(p_{\rm e}-\epsilon_{\rm e})}{w},\\
r^2\mathcal V&=
 \frac{4\pi r^2[5\epsilon_{\rm e}+9p_{\rm e}
             +(\epsilon_{\rm e}+p_{\rm e})\chi_{\rm e}]-6}{w} -4\left(\frac{m+4\pi r^3p_{\rm e}}{rw}\right)^2.
\end{align}
\end{subequations}
This is the relativistic quadrupole equation for a perfect fluid
\cite{Hinderer2008} applied to the effective source. The regular solution
has $H\propto r^2$ and $y\to2$ at the center. Outside the star its two
independent solutions are proportional to the associated Legendre
functions $P_2^2(r/M_0-1)$ and $Q_2^2(r/M_0-1)$, representing the growing
applied field and decaying response.

A density discontinuity modifies the surface value of $y$. At an
interface $r_d$ with continuous effective pressure and an outward energy
density decrease $\Delta\epsilon_{\rm e}$, integration of the singular
EOS term in Eq.~\eqref{eq:th-tidal-ode} gives
\begin{equation}
 y(r_d^+)-y(r_d^-)=
 -\frac{4\pi r_d^3\Delta\epsilon_{\rm e}}
        {m(r_d)+4\pi r_d^3p_{\rm e}(r_d)}.
 \label{eq:th-density-jump}
\end{equation}
For an internal transition at nonzero pressure, the pressure contribution
in this denominator is required \cite{TakatsyKovacs2020}. For the adopted
vacuum surface prescription, $p_{\rm e}(R_0)$ is neglected and
$y_R=y(R_0^-)-4\pi R_0^3\epsilon_{\rm e}(R_0)/M_0$. This is a
correction evaluated in the limit of vanishing surface pressure.

Matching the applied and decaying solutions determines the Love number.
With $\mathscr C=M_0/R_0$ and
$B=2-y_R+2\mathscr C(y_R-1)$, the result is
\begin{subequations}\label{eq:th-k2}
\begin{align}
 k_2&=\frac{(8/5)\mathscr C^5(1-2\mathscr C)^2B}{\mathcal D_k},\\
\mathcal D_k&=2\mathscr C[6-3y_R+3\mathscr C(5y_R-8)] +4\mathscr C^3[13-11y_R+\mathscr C(3y_R-2)
                         +2\mathscr C^2(1+y_R)]\nonumber\\
&\quad+3(1-2\mathscr C)^2B\ln(1-2\mathscr C).
\end{align}
\end{subequations}
We use the corrected static expression \cite{Hinderer2008,Hinderer2009Erratum}.
In the limit of weak gravity it reduces to
$k_2^{\rm N}=(2-y_R)/[2(y_R+3)]$.
The spherical value in Eq.~\eqref{eq:th-love-definitions} normalizes the
response at $\Omega=0$; it cannot be extended to a
rotating star by replacing $R_0$ with $R_{\rm eq}$.

\subsubsection{Coupling between rotation and tides}

Rotation and tide are expanded independently. Introduce formal
bookkeeping variables $s$ and $\tau$ for spin and tide, retain
$s^i\tau^j$ with $i\leq2$ and $j\leq1$, and set them to unity after
extracting coefficients. They carry no physical units; the fields
already include their physical factors of $\Omega$ and $\mathcal E$.
The product of the spin dipole with a polar quadrupole excites axial
$\ell=1,3$ perturbations at $O(\mathcal E\Omega)$; at
$O(\mathcal E\Omega^2)$ the polar sectors are $\ell=0,2,4$
\cite{Pani2015}. Subscripts s, t, and st below identify the isolated
spin at second order, static tide, and mixed polar fields. The complete
ansatz within Eq.~\eqref{eq:th-circular-metric} is
\begin{subequations}\label{eq:th-hierarchy}
\begin{align}
 h={}&s^2(h_{{\rm s}0}+h_{{\rm s}2}P_2)+\tau h_{{\rm t}2}P_2
 +s^2\tau\sum_{\ell=0,2,4}h_{{\rm st}\ell}P_\ell,\\
\mathfrak m={}&s^2(m_{{\rm s}0}+m_{{\rm s}2}P_2)+\tau m_{{\rm t}2}P_2
 +s^2\tau\sum_{\ell=0,2,4}m_{{\rm st}\ell}P_\ell,\\
k={}&s^2 k_{{\rm s}2}P_2+\tau k_{{\rm t}2}P_2
 +s^2\tau\sum_{\ell=2,4}k_{{\rm st}\ell}P_\ell,\\
\omega={}&s\omega_{\rm s}
       +s\tau[\omega_{{\rm t}1}P_1'(\mu)
                       +\omega_{{\rm t}3}P_3'(\mu)],\\
\Omega={}&s\Omega_{\rm s}+s\tau\delta\Omega_1.
\end{align}
\end{subequations}
All $P_\ell$ in this expression have argument $\mu$; only the primes on
these polynomials denote derivatives with respect to $\mu$.
The uniform shift $\delta\Omega_1$ is required by the constraint on
angular momentum. There is no monopole in $k$, including at mixed order.
One must expand the full metric, its inverse, and the normalized fluid
velocity. Retaining only an additive sum of linear metric perturbations
would discard products that contribute at $s^2\tau$.

Material perturbations follow from the effective Euler equation without
an independently imposed pressure ansatz. Define the dimensionless
enthalpy $W_{\rm e}$ by
$\dd W_{\rm e}=\dd p_{\rm e}/(\epsilon_{\rm e}+p_{\rm e})$.
For uniform circular flow its spatial first integral is
$W_{\rm e}-\ln u^t=C_{\rm E}$, where $C_{\rm E}$ is constant within
each equilibrium. Normalization gives
\begin{subequations}\label{eq:th-enthalpy-integral}
\begin{align}
 u^t&=\mathcal N^{-1/2},\\
\mathcal N&=-(g_{tt}+2\Omega g_{t\phi}+\Omega^2g_{\phi\phi}),\\
\delta W_{\rm e}&=\delta C_{\rm E}
                  -\frac12\ln\!\left(\frac{\mathcal N}{e^\nu}\right).
\end{align}
\end{subequations}
The integration constants contain a spin monopole shift, a possible
static monopole shift, and a mixed monopole shift $\delta C_2$.
A pure static quadrupole has zero monopole shift.
For any scalar $X(p_{\rm e})$, Taylor expansion of this first integral
implies
\begin{subequations}\label{eq:th-material-expansion}
\begin{align}
 X&=X_0+X_W\delta W_{\rm e}
                    +\tfrac12 X_{WW}(\delta W_{\rm e})^2,\\
X_W&=\mathcal H\frac{\dd X}{\dd p_{\rm e}},\\
X_{WW}&=\mathcal H^2\frac{\dd^2X}{\dd p_{\rm e}^2}
          +\mathcal H(1+\chi_{\rm e})\frac{\dd X}{\dd p_{\rm e}},\\
\mathcal H&=\epsilon_{\rm e}+p_{\rm e}.
\end{align}
\end{subequations}
Applying this expression to $X=p_{\rm e},\epsilon_{\rm e},n_B$ fixes all
fluid coefficients. In particular, if $W_{\rm s}$, $W_{\rm t}$, and
$W_{\rm st}$ are the $s^2$, $\tau$, and $s^2\tau$ coefficients of
$\delta W_{\rm e}$, then
$X_{\rm st}=X_W W_{\rm st}+X_{WW}W_{\rm s}W_{\rm t}$.
The second pressure derivatives in
Eqs.~\eqref{eq:th-eos-derivatives}--\eqref{eq:th-baryon-derivatives}
therefore enter the actual mixed equations.

Projection of the Einstein residual
$\mathscr R_{\alpha\beta}=G_{\alpha\beta}-8\pi T^{\rm e}_{\alpha\beta}$
yields the radial mixed system.
The expansion of this residual determines the nonlinear sources.
Insert the already determined spin, static tidal, and axial
fields into Eqs.~\eqref{eq:th-circular-metric}, \eqref{eq:th-hierarchy},
and \eqref{eq:th-material-expansion}, set the unknown mixed polar fields
and $\delta C_2$ to zero, and extract the $s^2\tau$ residual.
For a scalar function $f(\mu)$ define
$\Pi_\ell[f]=(2\ell+1)\int_{-1}^{1}P_\ell f\dd\mu/2$.
Let $\mathscr R_\ell=\Pi_\ell[\mathscr R_{rr}]$,
$\mathscr T_0=\Pi_0[\mathscr R^t{}_t]$, and define the coefficients of
the vector and traceless angular sources by
\begin{subequations}\label{eq:th-source-projections}
\begin{align}
 \mathscr V_\ell&=\frac{2\ell+1}{2\ell(\ell+1)}
       \int_{-1}^{1}(1-\mu^2)P_\ell'\mathscr R_{r\mu}\,\dd\mu,\\
\mathscr U_\ell&=\frac{(2\ell+1)(\ell-2)!}{2(\ell+2)!}
       \int_{-1}^{1}(1-\mu^2)P_\ell''
       \left(\frac{\mathscr R_{\mu\mu}}{g_{\mu\mu}}
                         -\frac{\mathscr R_{\phi\phi}}{g_{\phi\phi}}\right)
                         \dd\mu,\qquad \ell=2,4.
\end{align}
\end{subequations}
All quantities inside these projections denote the extracted residual
at the specified order, with metric contractions expanded consistently.
Writing $v_{{\rm st}\ell}=h_{{\rm st}\ell}+k_{{\rm st}\ell}$, the
inhomogeneous polar equations are
\begin{subequations}\label{eq:th-mixed-polar}
\begin{align}
 m_{{\rm st}\ell}&=-rw h_{{\rm st}\ell}+f_{m\ell},\\
v_{{\rm st}\ell}'&=-\nu'h_{{\rm st}\ell}+f_{v\ell},\\
h_{{\rm st}\ell}'&=\mathcal A h_{{\rm st}\ell}
                      +\mathcal B_\ell v_{{\rm st}\ell}+f_{h\ell},\\
f_{m\ell}&=r^3w\mathscr U_\ell,\\
f_{v\ell}&=(r+\nu'r^2/2)\mathscr U_\ell+\mathscr V_\ell,\\
f_{h\ell}&=\frac{\nu'r^2}{2}\mathscr U_\ell
        +\left(1+\frac{2}{r\nu'}\right)\mathscr V_\ell
        +\frac{\mathscr R_\ell}{\nu'}.
\end{align}
\end{subequations}
The homogeneous coefficients are those of
Eq.~\eqref{eq:th-polar-coefficients}; rotation enters the mixed polar
problem through the sources and boundary data.

At order $s\tau$, define $f_{{\rm ax}\ell}$ as minus the projected known
$t$--$\phi$ residual divided by the coefficient of
$\omega_{{\rm t}\ell}''$ in that equation. The axial radial equations are
\begin{equation}
 \omega_{{\rm t}\ell}''={}
 \left[\frac{\ell(\ell+1)-2}{r^2w}+\frac{16\pi\mathcal H}{w}\right]
                  \omega_{{\rm t}\ell}
 +\left(-\frac4r+\frac{4\pi r\mathcal H}{w}\right)
                  \omega_{{\rm t}\ell}' 
 -\delta_{\ell1}\frac{16\pi\mathcal H}{w}\delta\Omega_1
                       +f_{{\rm ax}\ell},\qquad \ell=1,3.
\label{eq:th-mixed-axial}
\end{equation}
Here $\delta_{\ell1}$ is the Kronecker symbol. These fields carry the
current multipoles induced by the electric tide in a spinning star.
Their feedback is required in the $s^2\tau$ sources, even when the final
quantity is an electric mass quadrupole.

The mixed monopole enters the constraint on physical baryon number. Define
$\mathcal A_0=4\pi r\mathcal H/w$,
$\mathcal B_0=(1+8\pi r^2p_{\rm e})/(r^2w^2)$, and
$\mathcal D_0=4\pi r^2\chi_{\rm e}\mathcal H$. The remaining equations are
\begin{subequations}\label{eq:th-mixed-monopole}
\begin{align}
 h_{{\rm st}0}'&=-\mathcal A_0h_{{\rm st}0}
                  +\mathcal B_0m_{{\rm st}0}
                  +\mathcal A_0\delta C_2-\frac r2\mathscr R_0,\\
m_{{\rm st}0}'&=-\mathcal D_0h_{{\rm st}0}
                  +\mathcal D_0\delta C_2+\frac{r^2}{2}\mathscr T_0.
\end{align}
\end{subequations}
Together with the residual prescription, these equations determine the
response from the coupled gravitational and fluid perturbations.

\subsubsection{Constraints, surface matching, and induced multipoles}

Regularity removes singular interior solutions. The homogeneous polar
solutions have $h_\ell\propto r^\ell$ and
$v_\ell\propto r^{\ell+2}$ for $\ell\geq2$, axial homogeneous solutions
have $\omega_\ell\propto r^{\ell-1}$, and a regular monopole has finite
$h_0$ and $m_0=O(r^3)$. Particular mixed solutions must be expanded
with their sources: products of fields at preceding orders can give lower powers
than these homogeneous expressions. For example, the full mixed
$v_{{\rm st}2}$ can contain an $O(r^2)$ particular term. Imposing an
$O(r^4)$ condition on that full field would remove part of the forced
solution.

Interpolation of the EOS also enters this matching problem. In the tidal
extension, mapped table nodes are interpolated linearly in logarithms,
so that within interval $i$
$\epsilon_{\rm e}=K_i p_{\rm e}^{\alpha_i}$ and
$n_B=N_i p_{\rm e}^{\beta_i}$, where $K_i,N_i,\alpha_i,\beta_i$ are
fixed by neighboring nodes. Consequently
$\chi_{\rm e}=\alpha_i\epsilon_{\rm e}/p_{\rm e}$ and
$\dd\chi_{\rm e}/\dd p_{\rm e}
 =\alpha_i(\alpha_i-1)\epsilon_{\rm e}/p_{\rm e}^2$ in that interval.
Although the interpolated energy and number densities are continuous,
their first pressure derivatives can jump. In a distributional
notation,
\begin{equation}
 \frac{\dd^2\epsilon_{\rm e}}{\dd p_{\rm e}^2}
 =\left(\frac{\dd\chi_{\rm e}}{\dd p_{\rm e}}\right)_{\rm bulk}
    +\sum_k[\chi_{\rm e}]_{p,k}\,
                     \delta(p_{\rm e}-p_{{\rm e}k}),
 \label{eq:th-eos-impulses}
\end{equation}
where $[\chi_{\rm e}]_{p,k}$ is the derivative on the side of higher
pressure minus that on the side of lower pressure at knot $k$.
The Dirac distribution here describes
a derivative discontinuity of the interpolant, not a physical density
transition. Its integral in the mixed equations, and the analogous
term for $n_B$, must be included along with the bulk solutions on either
side of each knot.

Both rotation and tide displace the material surface. Write
$R_\Sigma(\mu)=R_0+s^2\xi_{\rm s}(\mu)+\tau\xi_{\rm t}(\mu)
 +s^2\tau\xi_{\rm st}(\mu)$. Taylor expansion of the same surface
isobar gives
\begin{subequations}\label{eq:th-mixed-surface}
\begin{gather}
 \delta p_{{\rm e},{\rm s}}+p_{\rm e}'\xi_{\rm s}=0,\\
\delta p_{{\rm e},{\rm t}}+p_{\rm e}'\xi_{\rm t}=0,\\
\delta p_{{\rm e},{\rm st}}+p_{\rm e}'\xi_{\rm st}
 +\xi_{\rm s}\delta p_{{\rm e},{\rm t}}'
 +\xi_{\rm t}\delta p_{{\rm e},{\rm s}}'
 +p_{\rm e}''\xi_{\rm s}\xi_{\rm t}=0.
\end{gather}
\end{subequations}
All fields and radial derivatives in this equation are evaluated at
$R_0$. The expansion fixes displacement of the operational isobar;
setting each Eulerian pressure perturbation separately to zero at
$R_0$ would omit the moving boundary.

Junction conditions are imposed on that displaced hypersurface.
Let $e_A^\alpha$ be its tangent vectors and $n_\alpha$ its outward unit
normal. With coordinates $(t,r,\mu,\phi)$,
\begin{subequations}\label{eq:th-junction-geometry}
\begin{align}
 n_\alpha&=\frac{(0,1,-\dd R_\Sigma/\dd\mu,0)}
 {\sqrt{g^{rr}+g^{\mu\mu}(\dd R_\Sigma/\dd\mu)^2}},\\
q_{AB}&=g_{\alpha\beta}e_A^\alpha e_B^\beta,\\
K_{AB}&=e_A^\alpha e_B^\beta\nabla_\alpha n_\beta.
\end{align}
The indices $A,B$ label coordinates intrinsic to the hypersurface.
Matching without a material thin shell requires
$[q_{AB}]=[K_{AB}]=0$, with brackets denoting exterior minus interior
values. We expand these equations through the same spin and tide orders,
using the surface prescription in which the residual pressure is neglected.
At second order the displacement contributes to the junction conditions;
continuity of each radial perturbation at the original radius is not
sufficient
\cite{ReinaVera2015}.

A spatial surface visualization follows from the same induced geometry.
On the declared slice of constant time, set
$\mathcal R_\Sigma(\theta)=R_\Sigma(\cos\theta)$ and define
$\gamma_{\theta\theta}
 =[g_{\theta\theta}+g_{rr}(\dd\mathcal R_\Sigma/\dd\theta)^2]_{r=\mathcal R_\Sigma}$
and $\gamma_{\phi\phi}=[g_{\phi\phi}]_{r=\mathcal R_\Sigma}$.
The product involving surface derivatives is retained when it contributes at mixed
order. An isometric embedding as a Euclidean surface of revolution,
with cylindrical radius $X$ and height $Z_{\rm emb}$, is obtained from
\begin{align}
 \dd\ell_{\rm surf}^2&=\gamma_{\theta\theta}\dd\theta^2
                         +\gamma_{\phi\phi}\dd\phi^2,
                         \label{eq:th-surface-two-metric}\\
 X(\theta)&=\sqrt{\gamma_{\phi\phi}(\theta)},\\
 \frac{\dd Z_{\rm emb}}{\dd\theta}
   &=-\sqrt{\gamma_{\theta\theta}
                 -\left(\frac{\dd X}{\dd\theta}\right)^2},
                 \qquad Z_{\rm emb}(\pi/2)=0.
                 \label{eq:th-surface-embedding}
\end{align}
Here $\dd\ell_{\rm surf}$ is proper distance on the spatial surface.
The sign specifies the northern meridian, and equatorial symmetry
completes the contour. All products and square roots are expanded to
retain $1,\Omega^2,\mathcal E,\mathcal E\Omega^2$, consistently with
the stellar solution. At the equator $X$ is the circumferential radius;
$Z_{\rm emb}(0)$ is an embedded polar height and is not generally a coordinate
polar radius. This construction uses the displaced isobar and its
angular metric, so the surface contour is not inferred from the induced
mass quadrupole alone.
\end{subequations}

Conservation of the charges in Eq.~\eqref{eq:th-charges} supplies the
remaining fluid constants. At $O(\mathcal E\Omega)$ the condition
$\delta J=0$ fixes the uniform $\delta\Omega_1$ together with the axial
matching constants. At $O(\mathcal E\Omega^2)$, $\delta A=0$ fixes the
monopole shift $\delta C_2$ after the coupled surface and exterior
conditions have been applied. The constraints include the moving
integration domain. In particular, for any radial integrand expanded as
$F=F_0+s^2F_{\rm s}+\tau F_{\rm t}+s^2\tau F_{\rm st}$, the mixed
coefficient of its integral to $R_\Sigma$ is
\begin{equation}
 \left[\int_0^{R_\Sigma} F\,\dd r\right]_{s^2\tau}
 =\int_0^{R_0}F_{\rm st}\,\dd r
 +F_0\xi_{\rm st}+F_{\rm s}\xi_{\rm t}+F_{\rm t}\xi_{\rm s}
 +F_0'\xi_{\rm s}\xi_{\rm t},
 \label{eq:th-moving-charge}
\end{equation}
where the nonintegrated terms are evaluated at $R_0$ before angular
integration. These terms and those at the EOS knots implement conservation
of physical baryons without introducing an effective particle count.

Outside the star, the solution contains the applied electric field and
induced mass and current multipoles. We adopt the canonical applied/response
split of the exterior expanded through second order in rotation
\cite{Pani2015}. The full applied
solution, including the terms that depend on spin, is retained in surface
matching. Only for extracting the induced moment is the complete
canonical applied mode subtracted, leaving an asymptotically flat
response metric. Subtracting only the leading growing $r^2P_2$ term
would leave contributions from the applied field in the inferred
response at finite spin.

To extract the electric mass quadrupole from the vacuum metric, we use
the invariant expansion of circular geodesics \cite{Ryan1995}. To specify
the normalization, write the equatorial asymptotic response metric as
\begin{subequations}\label{eq:th-asymptotic-metric}
\begin{align}
 g_{tt}&=-1+\frac{2\mathcal M}{r}+\frac{a_2}{r^2}
                         +\frac{a_3}{r^3}+O(r^{-4}),\\
g_{\phi\phi}&=r^2+b_1r+b_0+O(r^{-1}).
\end{align}
\end{subequations}
Here $\mathcal M$ and the coefficients are formal series through the
retained orders; $[a_2]=[b_0]=L^2$, $[a_3]=L^3$, and $[b_1]=L$.
Expansion of the angular velocity and specific energy of circular orbits
in the invariant orbital variable gives the mass moment
\begin{equation}
 M_2=\frac{
 4a_2^2-4a_2b_1\mathcal M-18a_2\mathcal M^2 
 {}-12a_3\mathcal M+b_1^2\mathcal M^2
 -18b_1\mathcal M^3-12b_0\mathcal M^2
 }{12\mathcal M}.
 \label{eq:th-ryan-moment}
\end{equation}
All products and the division in this equation are expanded to the
retained perturbative orders. The corresponding vacuum relation
$a_2+b_1\mathcal M=0$ provides a consistency check. The Cartesian component of the quadrupole with zero trace is $Q_{zz}=2M_2/3$, which fixes the normalization of
the axisymmetric tidal coefficient. Removing the isolated rotational
moment, we define the response of each equilibrium by
\begin{equation}
 \lambda_{\rm T}(A,J)
 =-\left.\frac{\partial Q_{zz}^{\rm ind}}{\partial\mathcal E}
                                                     \right|_{A,J} 
 =\lambda_{\rm static}(p_c)
       +\delta\lambda_{\rm mixed}(p_c,\Omega)+O(\Omega^4).
\label{eq:th-rotating-love}
\end{equation}
For the selected stationary aligned electric component, the response
coefficient is $\lambda_{{\rm T},20}$. In a rotating star the full tidal response
has additional components. Holding $A,J$ fixed specifies the adjustment
of the uniformly rotating fluid; a dynamical tide with a prescribed
circulation requires conditions on the fluid motion beyond those imposed
here. Comparisons of rotational tidal coefficients must use the same
convention for separating the applied field and the response
\cite{Pani2015}.

If a dimensionless response is normalized by the rotating mass, that
normalization must also be expanded. With $M=M_0+\delta M$, it is
\begin{equation}
 \Lambda_{\rm T}=\frac{\lambda_{\rm static}}{M_0^5}
 +\frac{\delta\lambda_{\rm mixed}}{M_0^5}
 -\frac{5\delta M\lambda_{\rm static}}{M_0^6}+O(\Omega^4).
 \label{eq:th-rotating-normalization}
\end{equation}
Normalizing by the rotating mass produces the last term. It must be
distinguished from the change in the dimensional induced quadrupole when
interpreting the rotational correction.

\subsection{Structural and mixed contributions at fixed baryon number}\label{sec:decomposition}

Along an isolated sequence at fixed $A$, rotation changes the central
pressure from its initial static value $p_{c,0}$ to $p_{c,\Omega}$.
The associated spherical background is labeled by the same
$p_{c,\Omega}$; its baryon number need not equal the prescribed $A$
of the rotating equilibrium. At each coupling, define
$\lambda_{\rm ref}=\lambda_{\rm T}(A,0)
                  =\lambda_{\rm static}(p_{c,0})$.
The computed rotating response can be separated as
\begin{subequations}\label{eq:th-decomposition}
\begin{align}
 \lambda_{\rm T}(A,x)&=
 \lambda_{\rm static}(p_{c,\Omega})
 +\delta\lambda_{\rm mixed}(p_{c,\Omega},\Omega),\\
S&=\frac{\lambda_{\rm static}(p_{c,\Omega})}{\lambda_{\rm ref}}-1,\\
C&=\frac{\delta\lambda_{\rm mixed}(p_{c,\Omega},\Omega)}
                                                 {\lambda_{\rm ref}},\\
\delta_{\rm rot}\lambda_{\rm T}
 &\equiv\frac{\lambda_{\rm T}(A,x)}{\lambda_{\rm ref}}-1=S+C.
\end{align}
\end{subequations}
For the retained numerical responses, the sum is an algebraic identity.
The structural term $S$ compares two spherical backgrounds, while $C$
contains the explicit mixed solution on the background associated with
the rotating equilibrium. This separation depends on the background
convention and does not define independently measurable tidal
components. Evaluating the functions at $p_{c,\Omega}$ includes the
background adjustment along the sequence. It does not determine the
omitted terms beyond second order in rotation about $p_{c,0}$.

Using the definition of the spherical Love number at both central pressures
makes the structural content explicit:
\begin{equation}
 1+S=\frac{k_2(p_{c,\Omega})}{k_2(p_{c,0})}
       \left[\frac{R_0(p_{c,\Omega})}{R_0(p_{c,0})}\right]^5.
 \label{eq:th-structural-identity}
\end{equation}
Changes in the length scale enter through the radius factor, while the
Love number ratio measures the change of the dimensionless internal response.
Both radii belong to spherical reference models;
$R_0(p_{c,\Omega})$ is distinct from the equatorial radius of the
associated rotating configuration. This identity connects the background
readjustment along a sequence at fixed $A$ to the structural contribution
$S$ in Eq.~\eqref{eq:th-decomposition}.

\section{Results and Discussion}\label{sec:results}

The equilibrium sequences and tidal responses are calculated for cold
$^{12}\mathrm{C}$ matter with the same electron, ionic, and lattice
contributions to the equation of state (EOS). We consider
$\lambda=0,-5\times10^{-5},-10^{-4},-2\times10^{-4},-3\times10^{-4}$,
and $-4\times10^{-4}$ in $f(R,T)=R+2\lambda T$, with general relativity
(GR) recovered at $\lambda=0$. The spherical reference models have
radius $R_0\leq12000\ \mathrm{km}$ and gravitational mass $M_0$.
The sampled central densities satisfy
$\rho_c\leq9.26\times10^9\ \mathrm{g\,cm^{-3}}$, below the reference
carbon capture threshold. This density selection and the radius cutoff
define the range represented in the figures.
Rotation is parametrized by $x=\Omega/\Omega_K$, where $\Omega_K$ is the
Kepler reference evaluated for each model. Thus, equal values of $x$
do not imply equal angular velocities across masses or couplings.
The distinction between sequences at prescribed central density and
sequences at fixed physical baryon number $A$ will be essential when
relating changes in global structure to the tidal response.

\begin{figure}[!htbp]
\centering
\includegraphics[width=0.60\textwidth]{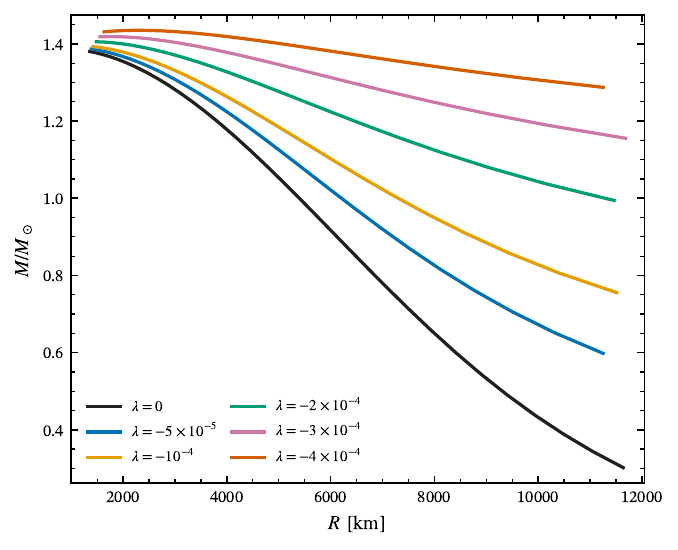}
\caption{Static mass--radius relations for cold $^{12}\mathrm{C}$
fluid white dwarfs in $f(R,T)=R+2\lambda T$. The legend identifies the
six couplings, including the GR limit $\lambda=0$.
Mass is measured in $M_\odot$ and radius in km. All curves use the same
EOS and operational surface prescription, with $R_0\leq12000\ \mathrm{km}$.
The endpoints delimit the selected central densities and are not
uniformly determined stability limits.}
\label{fig:static-mr}
\end{figure}

Without rotation, a more negative coupling shifts the mass--radius
relation toward larger radii over the common mass interval
(Fig.~\ref{fig:static-mr}). The largest masses sampled increase from
$1.380\,M_\odot$ in GR to $1.435\,M_\odot$ at
$\lambda=-4\times10^{-4}$, a change of about $4.0\%$. These numbers
characterize the selected density interval: for several couplings the
largest mass occurs at its upper endpoint, so they do not all identify
resolved extrema of complete equilibrium sequences. The different
low-mass endpoints also reflect the radius selection and should not
be interpreted as physical minimum masses.

The effective pressure defined in Eq.~\eqref{eq:th-map} explains the
modified radial balance. For negative coupling, the contribution
proportional to $\mathrm{d}\epsilon/\mathrm{d}p$ increases
$D=\mathrm{d}p_{\rm e}/\mathrm{d}p$. This sensitivity is strongest in
layers where $\mathrm{d}p/\mathrm{d}\epsilon$ is small, allowing a
small coupling to affect the extended stellar layers while the
material EOS remains fixed. The larger radii and modest mass increase
agree qualitatively with Carvalho et al.~\cite{Carvalho2017}.
Quantitative comparison requires the same EOS, density range, and
surface prescription, since the outer layers are sensitive to the
location of the adopted surface.

\begin{figure}[!htbp]
\centering
\includegraphics[width=0.99\textwidth]{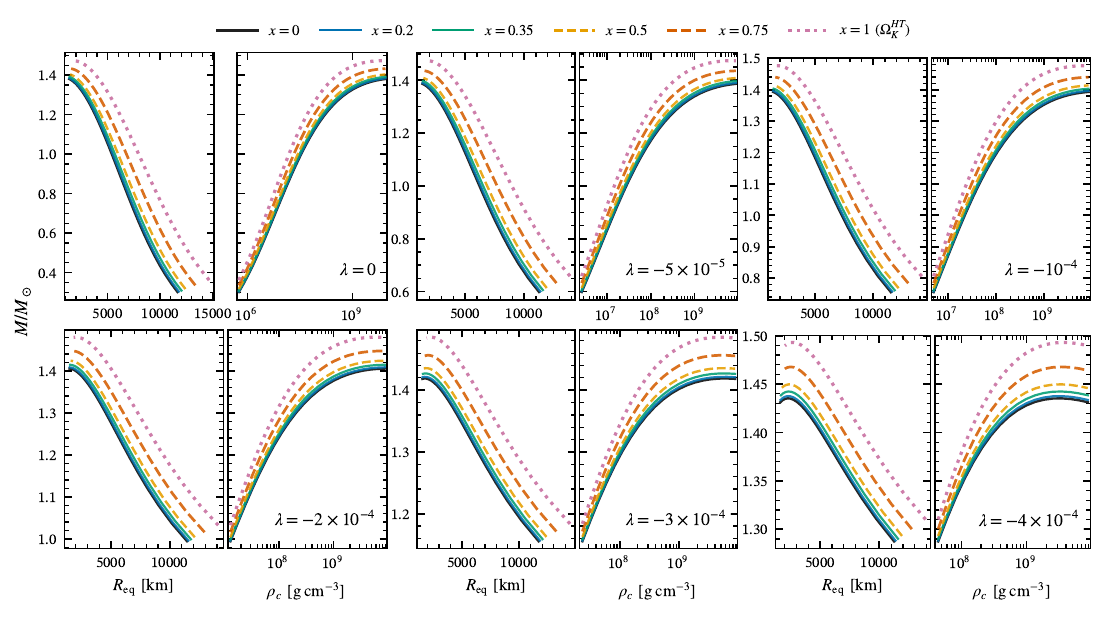}
\caption{Rotating equilibrium sequences for the six couplings indicated
inside the panels. Each group combines $M$ versus equatorial radius
$R_{\rm eq}$ (km) with $M$ versus central density $\rho_c$
($\mathrm{g\,cm^{-3}}$), sharing the mass axis in $M_\odot$.
The legend gives $x=\Omega/\Omega_K=0,0.20,0.35,0.50,0.75,1$.
Different spins are evaluated on the same central-density grid for each
coupling, with $R_0\leq12000\ \mathrm{km}$ for the spherical references.
Solid curves have $x\leq0.35$; dashed and dotted curves extend the
second-order expansion to higher spin. The Kepler reference and axis
ranges are model dependent.}
\label{fig:rotating-structure}
\end{figure}

Adding rotation changes both the mass and the spatial extent of these
equilibria, as shown by the paired mass--equatorial-radius and
mass--central-density relations in Fig.~\ref{fig:rotating-structure}.

At a given coupling and central density, increasing $x$ raises the
calculated gravitational mass and equatorial radius. The central-density
representation identifies which configurations share the same spherical
reference, whereas the radius representation shows the resulting change
in size. A displacement in the mass--radius plane can therefore receive
contributions from both the matter--geometry coupling and rotation.
This overlap of theoretical effects does not by itself establish an
observational degeneracy, since the relevant measurement errors and
uncertainties in the stellar composition have not been included.

Across the structural grid, the largest
$\epsilon_{\rm rot}=\Omega^2R_0^3/M_0$ is approximately $0.077$ at
$x=0.35$, $0.158$ at $x=0.50$, and $0.631$ at $x=1$.
Quantitative comparisons use $\epsilon_{\rm rot}\leq0.1$; curves
extending to each model's Kepler reference illustrate the retained
second order solution without controlling higher orders in rotation.
The radius selection applies to $R_0$, allowing
$R_{\rm eq}>12000\ \mathrm{km}$ at nonzero spin.

\begin{figure}[!htbp]
\centering
\includegraphics[width=0.99\textwidth]{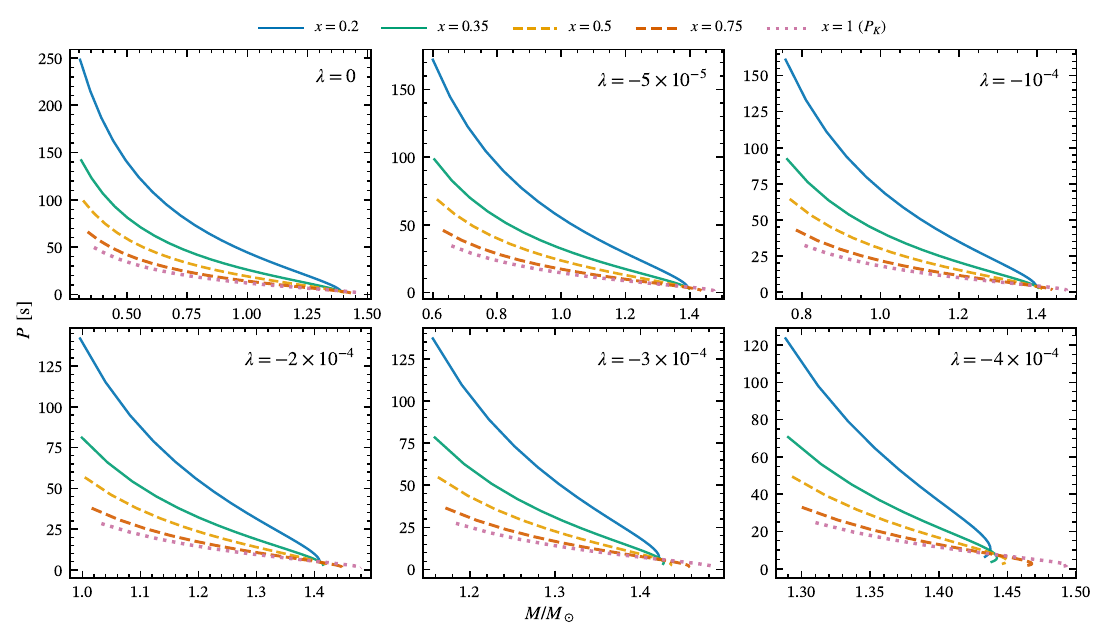}
\caption{Rotation period $P_{\rm rot}=2\pi/\Omega$ in seconds versus
rotating gravitational mass $M$ in $M_\odot$.
Each panel has the coupling shown internally; the legend identifies
$x=\Omega/\Omega_K=0.20,0.35,0.50,0.75,1$, with $x=1$ giving the
reference $P_K$. The static case is absent because its period is
infinite. Curves follow the selected central-density sequences.
Solid curves correspond to $x=0.20,0.35$; the higher-spin curves
continue the second-order solution beyond the controlled slow-rotation
domain where its criteria are exceeded.}
\label{fig:period-mass}
\end{figure}

The periods in Fig.~\ref{fig:period-mass} translate the same rotation
parameter into a physical time scale,
$P_{\rm rot}=2\pi/(x\Omega_K)$.

For a given reference model, increasing
$x$ shortens the period, while the variation along a mass sequence
also reflects the changing Kepler frequency. The tendency toward
shorter periods for more massive, smaller configurations is consistent
with the leading dynamical scaling
$P_K\sim2\pi(R_0^3/M_0)^{1/2}$. This scaling is used only to interpret
the computed periods, which employ the actual declared Kepler
reference. Near the mass extrema, different central densities can
correspond to similar gravitational masses and different periods;
$P_{\rm rot}(M)$ is therefore not globally single valued.
The figure supplies equilibrium period scales for later comparison
with stellar measurements, without imposing an observational
constraint on $\lambda$ or identifying the shortest plotted period
with a demonstrated astrophysical limit.

\begin{figure}[!htbp]
\centering
\includegraphics[width=0.99\textwidth]{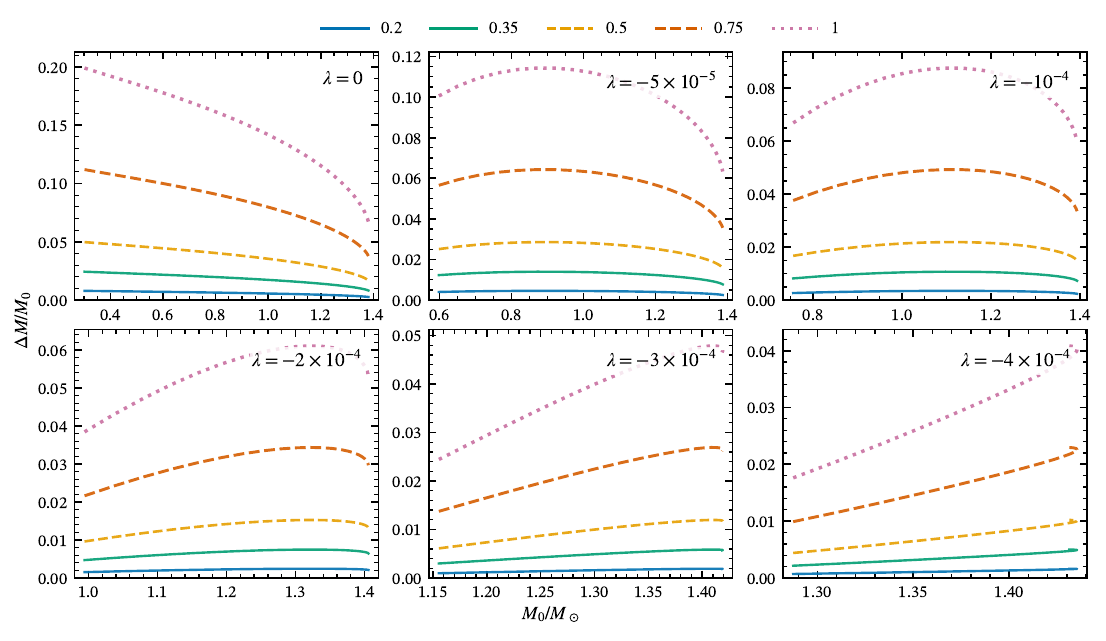}
\caption{Fractional rotational mass correction
$\Delta M/M_0=(M-M_0)/M_0$ versus spherical gravitational mass
$M_0/M_\odot$, at the same central density and coupling.
The six panels give the indicated values of $\lambda$ for the cold
$^{12}\mathrm{C}$ EOS. Legend values denote
$x=\Omega/\Omega_K=0.20,0.35,0.50,0.75,1$.
The $x=0$ correction vanishes. Solid curves show the two lower spins;
dashed and dotted curves show the higher-spin continuation of the
second-order expansion. This comparison does not hold $A$ fixed.}
\label{fig:delta-mass}
\end{figure}

To separate the rotational contribution from the change already present
in the static star, Fig.~\ref{fig:delta-mass} gives
$\Delta M/M_0=(M-M_0)/M_0$ at the same central density and coupling.

The correction is positive throughout the selected grid and increases
with spin. At $x=0.35$, it ranges from approximately $0.216\%$ to
$2.44\%$ across the six families. Its mass dependence changes with
$\lambda$: the GR correction decreases with mass, while the modified
sequences develop broad interior maxima or predominantly increasing
responses over their displayed mass intervals. The same quadratic
order in $\Omega$ therefore does not imply the same dependence on
stellar mass. Both the response coefficients and the conversion from
$x$ to $\Omega$ depend on the equilibrium structure. Since $A$ changes
in this comparison, $\Delta M$ is the mass correction to a prescribed
spherical background and is not the mass difference obtained by
rotating an initial star while keeping its baryon number fixed.

\begin{figure}[!htbp]
\centering
\includegraphics[width=0.99\textwidth]{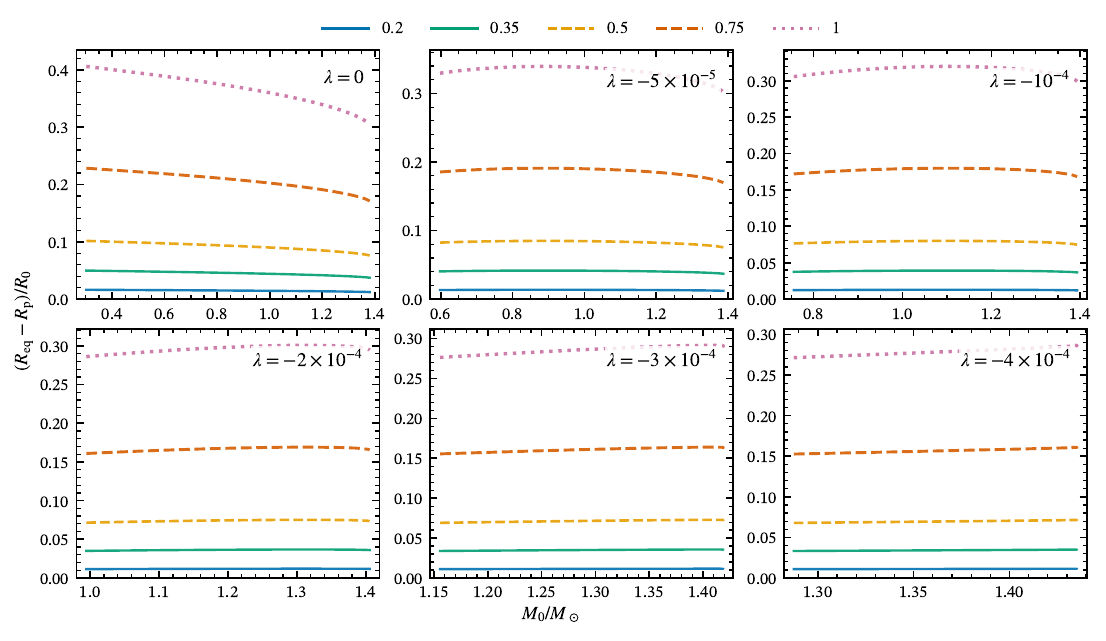}
\caption{Surface deformation
$\mathcal{D}=(R_{\rm eq}-R_{\rm p})/R_0$ versus $M_0/M_\odot$ for
the six couplings shown in the panels. Here $R_0$ is the spherical
reference radius and $R_{\rm p}$ is the polar radius; the central
density is matched between each rotating model and its reference.
Legend values are $x=\Omega/\Omega_K=0.20,0.35,0.50,0.75,1$.
Positive $\mathcal{D}$ denotes oblateness. Solid curves correspond
to $x=0.20,0.35$; dashed and dotted curves are the higher-spin
continuation of the second-order model.}
\label{fig:deformation}
\end{figure}

Surface distortion provides a complementary measure of the same
rotational adjustment. The positive quantity
$\mathcal{D}=(R_{\rm eq}-R_{\rm p})/R_0$ in
Fig.~\ref{fig:deformation} identifies oblate configurations, with
$R_{\rm eq}>R_{\rm p}$.

Its growth with spin follows the angular
dependence of centrifugal support, which vanishes on the rotation
axis and is largest in the equatorial plane. At $x=0.35$,
$\mathcal{D}$ lies between about $0.0333$ and $0.0497$ across the
sample. This relatively narrow range accompanies a more varied mass
dependence of $\Delta M/M_0$: the angular distortion of the surface
and the change in total mass need not track each other proportionally.
Here $\mathcal{D}$ is a radius difference normalized by $R_0$,
not the geometrical eccentricity
$[1-(R_{\rm p}/R_{\rm eq})^2]^{1/2}$.

\begin{figure}[!htbp]
\centering
\includegraphics[width=0.99\textwidth]{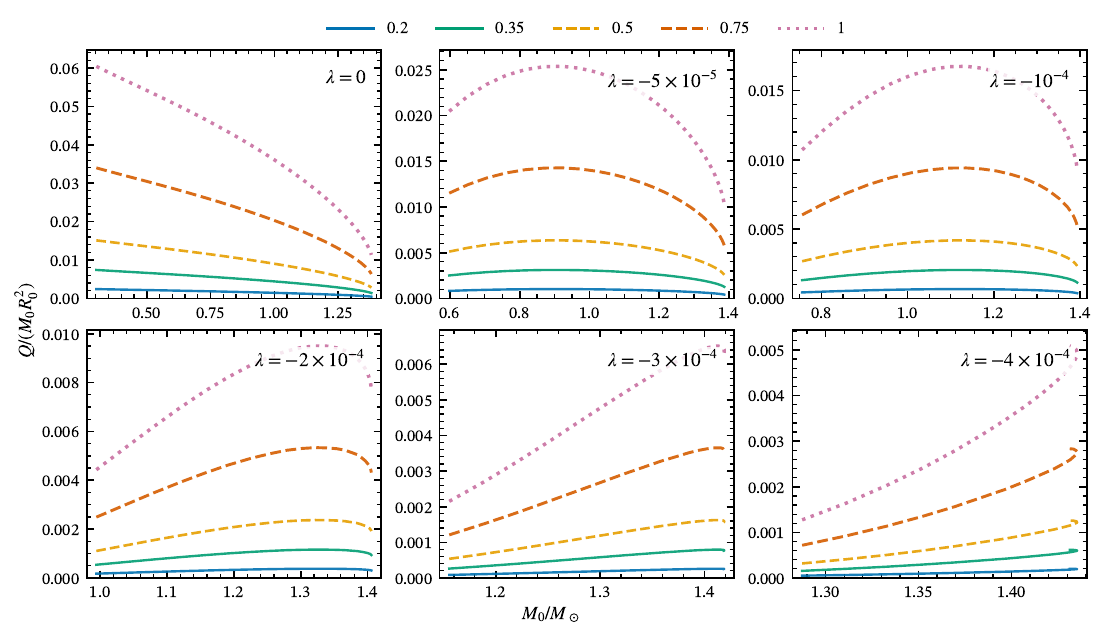}
\caption{Rotational quadrupole $Q/(M_0R_0^2)$ versus $M_0/M_\odot$
for the indicated gravitational couplings. The normalization is
dimensionless, with $M_0$ expressed as a geometrized length.
$Q$ follows the positive potential-coefficient convention for oblate
stars and is present in the absence of a tidal field.
Legend values specify $x=\Omega/\Omega_K$; central density is
matched to each spherical reference. Solid curves have
$x=0.20,0.35$, while the higher-spin curves continue the
second-order expansion.}
\label{fig:rotation-quadrupole}
\end{figure}

An external measure of the rotationally distorted mass distribution is
given by $Q/(M_0R_0^2)$ in Fig.~\ref{fig:rotation-quadrupole}.

At $x=0.35$ the displayed values span approximately
$1.56\times10^{-4}$ to $7.40\times10^{-3}$, a broader relative
variation than that of $\mathcal{D}$. A surface radius difference
does not determine how the entire interior contributes to the
quadrupole, and the normalization itself changes with mass and
coupling. The positions of the quadrupole maxima consequently need
not coincide with extrema of the surface deformation.
The positive $Q$ used here is the potential coefficient associated
with oblateness; the corresponding Geroch--Hansen mass moment
has the opposite sign in these conventions. This rotational
quadrupole exists without an applied tide and must be separated
from the induced quadrupole when extracting a tidal response.

\begin{figure}[!htbp]
\centering
\includegraphics[width=0.99\textwidth]{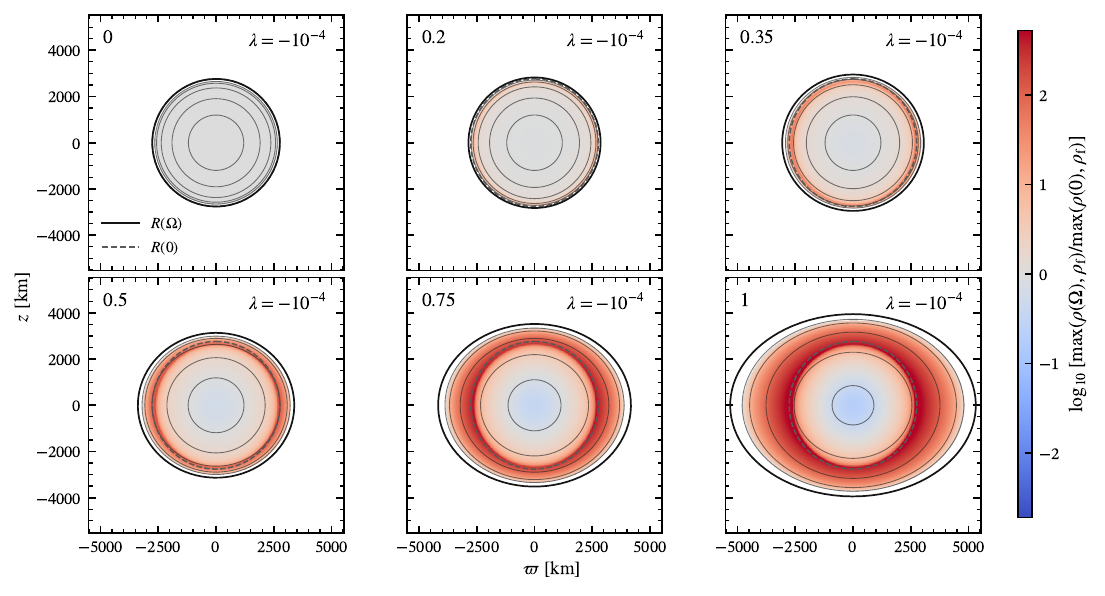}
\caption{Meridional density redistribution at $\lambda=-10^{-4}$
and fixed $\log_{10}A=57.20715121304262$. Panels give
$x=\Omega/\Omega_K=0,0.20,0.35,0.50,0.75,1$ in cylindrical
coordinates $(\varpi,z)$, in km. Color shows
$\log_{10}\{\max[\rho(x),\rho_{\rm f}]/\max[\rho(0),\rho_{\rm f}]\}$
at the same spatial position, with
$\rho_{\rm f}=10^4\ \mathrm{g\,cm^{-3}}$.
Thin contours mark available integer decades of density in
$\mathrm{g\,cm^{-3}}$; solid and dashed outlines are the rotating
and static surfaces. The map reconstructs nested isodensity shells
from their equatorial and polar radii. Panels with $x\geq0.50$
extend the second-order solution beyond the adopted slow-rotation
domain.}
\label{fig:density-redistribution}
\end{figure}

These comparisons at prescribed central density do not yet identify
the response of a fixed amount of stellar matter. Maintaining $A$
instead requires a change in central density as the rotational
support increases. Figure~\ref{fig:density-redistribution} illustrates
this adjustment at $\lambda=-10^{-4}$ for the sequence with
$\log_{10}A=57.20715121304262$.

Between $x=0$ and $x=0.35$, the
central density decreases from approximately
$8.36\times10^8$ to $6.32\times10^8\ \mathrm{g\,cm^{-3}}$, while
the equatorial radius increases from $2762$ to $3067\ \mathrm{km}$.
The polar radius also increases, reaching $2953\ \mathrm{km}$.
There is no inconsistency between polar expansion relative to the
initial star and oblateness of the rotating star: the overall
expansion and the equatorial--polar difference are distinct
contributions to the change in shape.

In this map, color compares densities at the same spatial position,
using a floor to define the ratio where one configuration has
negligible density. Positive values near the expanded surface
can therefore include positions outside the original star.
They indicate an extended density distribution, not a measured
fraction of matter deposited in an outer shell. The nested-shell
reconstruction supplies a spatial representation of the equilibrium
profiles; it does not calculate a time-dependent material flux or
an elastic crust. This distinction also matters for the tidal
calculation: keeping the physical baryon number fixed constrains
the amount of matter, while allowing its density profile and
gravitational mass to adjust. The same identification will be used to
separate this isolated rearrangement from the surface response to an
applied tide.

\begin{figure}[!htbp]
\centering
\includegraphics[width=0.99\textwidth]{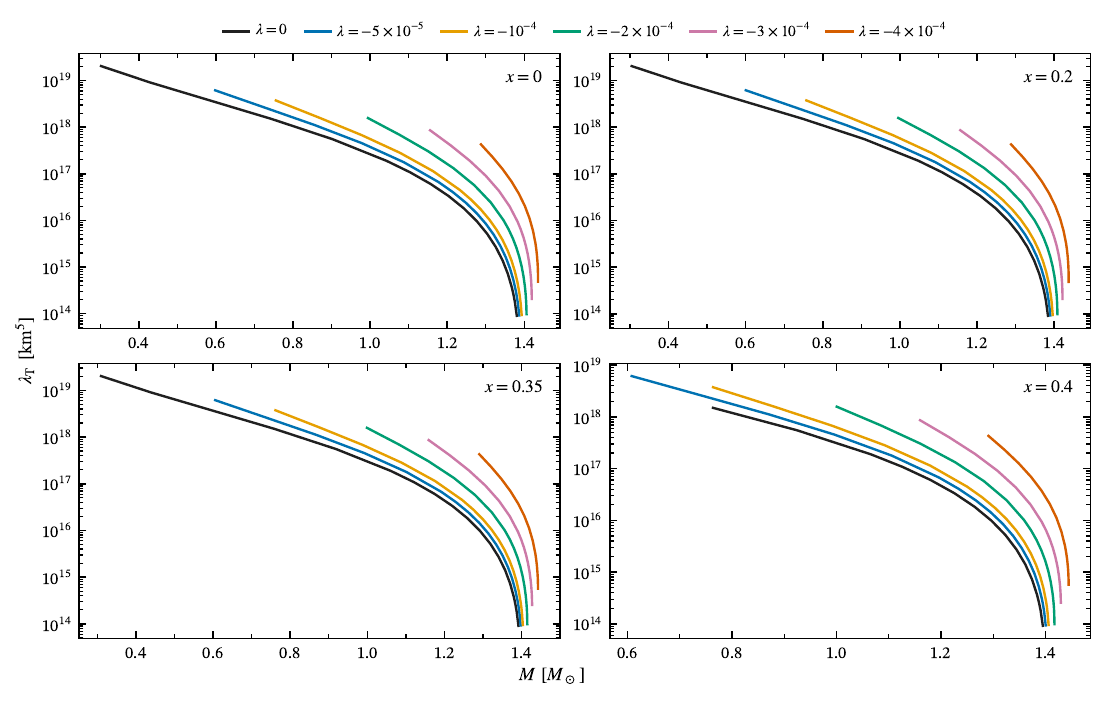}
\caption{Dimensional stationary electric tidal deformability
$\lambda_{\rm T}$ ($\mathrm{km}^5$) versus rotating gravitational
mass $M$ ($M_\odot$), for cold $^{12}\mathrm{C}$ fluid white
dwarfs. Panels give $x=\Omega/\Omega_K=0,0.20,0.35,0.40$;
colors identify $\lambda$ as in the legends.
The response is linear in the aligned $\ell=2,m=0$ tide and
retains rotation through second order, with $A,J$ fixed in each
tidal derivative. $A$ varies along the plotted mass sequences.
Only models satisfying the adopted domain and numerical controls
are shown; gaps are preserved. Shading gives empirical numerical
and surface-sensitivity envelopes, not statistical intervals or
bounds on omitted rotation orders.}
\label{fig:tidal-mass-rotation}
\end{figure}

An external tide probes this rearranged interior through the induced
mass quadrupole. We evaluate the aligned electric component of the
gravitational tide, $\ell=2,m=0$, at fixed $A,J$.
The normalization in Eq.~\eqref{eq:th-love-definitions} distinguishes
$\lambda_{\rm T}$, with units of $\mathrm{km}^5$, from the dimensionless
$\Lambda_{\rm T}$ \cite{BinningtonPoisson2009}.
Figure~\ref{fig:tidal-mass-rotation} shows the dimensional response.
At nonzero spin this includes the explicit mixed perturbations
introduced in Section~\ref{sec:tides}.

Across the mass sequences, $\lambda_{\rm T}$ decreases by several
orders of magnitude as the configurations become more massive
and smaller. At zero spin, Eq.~\eqref{eq:th-love-definitions}
separates the strong radius dependence from the dimensionless
response of the interior. The larger radii at negative coupling
are accompanied by larger dimensional deformabilities over the
common mass range, but this ordering does not require $k_2$
itself to increase. The response at nonzero spin retains the same
broad mass dependence. Its comparison between panels includes a
shift of the mass coordinate and a change in the baryon content,
so the visual proximity of the curves does not measure the
rotational change of an individual fixed-$A$ sequence.
The restricted low-mass coverage at $x=0.40$ follows the
model-by-model domain selection; it is not evidence for a new
physical minimum mass.

The vacuum analysis of Pani et al.~\cite{Pani2015} supplies the
multipolar coupling structure described in Section~\ref{sec:tides};
the stellar interior and its matching determine the white dwarf
response. Murshid and Kalam~\cite{MurshidKalam2024} studied rotation
and tidal deformability of neutron stars in the same gravity model,
but evaluated the tide for a nonspinning star. Their static result
therefore does not provide a quantitative benchmark for the mixed
rotational tidal terms calculated here.

\begin{figure}[!htbp]
\centering
\includegraphics[width=0.68\textwidth]{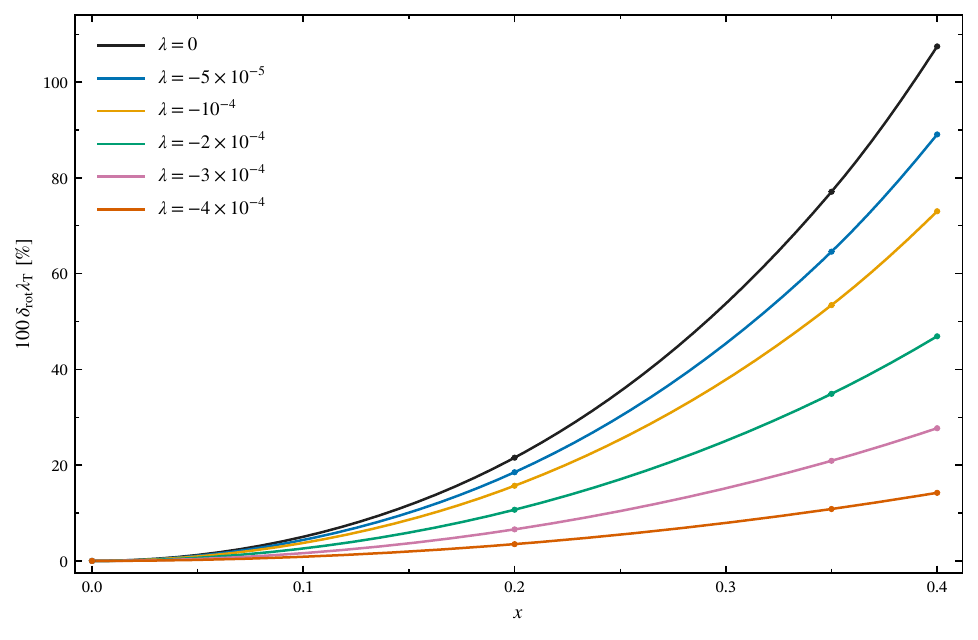}
\caption{Percentage change $100\delta_{\rm rot}\lambda_{\rm T}$
along the equilibrium sequence with
$\log_{10}A=57.20715121304262$, for the six couplings in the
legend. Each curve is normalized by its own static
$\lambda_{\rm T}(A,0)$ and uses its model-specific Kepler
reference. The response is the stationary aligned electric
$\ell=2,m=0$ component, with $A,J$ conserved during the tidal
derivative. Markers give the four calculated spins
$x=0,0.20,0.35,0.40$; lines are PCHIP interpolation in $x^2$
for visualization only. Bars show the empirical control
envelopes at the calculated nodes, mostly smaller than the
markers. $J$ can vary between equilibria.}
\label{fig:tidal-fixed-A}
\end{figure}

To follow the change at a fixed amount of matter,
Fig.~\ref{fig:tidal-fixed-A} uses the same baryon target as the
density reconstruction, now for all six couplings. The central
pressure $p_{c,\Omega}$ is adjusted at each spin to preserve $A$.
The plotted quantity, defined in Eq.~\eqref{eq:th-decomposition}, is
\begin{equation}
 \delta_{\rm rot}\lambda_{\rm T}
 =\frac{\lambda_{\rm T}(A,x)}{\lambda_{\rm T}(A,0)}-1.
 \label{eq:results-fixed-A-change}
\end{equation}
The baryon target is common to the isolated equilibria, whereas
$J$ varies with spin; $A$ and $J$ are both fixed during each
tidal derivative. Each coupling has its own static denominator. The response
increases monotonically at the calculated spins, while its
fractional increase becomes smaller for more negative $\lambda$.
At $x=0.40$, the increase is approximately $107.44\%$ in GR,
$73.02\%$ at $\lambda=-10^{-4}$, and $14.23\%$ at
$\lambda=-4\times10^{-4}$. Table~\ref{tab:tidal-fixed-A}
places these percentages beside the absolute deformabilities.
Although the fractional enhancement is smallest at
$\lambda=-4\times10^{-4}$, both its static and rotating
deformabilities are larger than the corresponding GR values.
The two trends answer different questions and are compatible.

\begin{table}[!htbp]
\caption{Tidal response at $\log_{10}A=57.20715121304262$
and $x=0.40$. The static and rotating dimensional
deformabilities are in units of $10^{15}\ \mathrm{km}^5$.
The last three columns give the total fractional change and
its structural and explicit mixed contributions, multiplied
by 100. All entries are evaluated at calculated nodes;
small discrepancies in their displayed sum arise from
rounding.}
\label{tab:tidal-fixed-A}
\begin{ruledtabular}
\begin{tabular}{crrrrrr}
$\lambda$ & $\lambda_{\rm T}(A,0)$ & $\lambda_{\rm T}(A,0.40)$
 & $100\delta_{\rm rot}\lambda_{\rm T}$ & $100S$ & $100C$\\
$0$                    & $1.314$ & $2.725$ & $107.44$ & $104.73$ & $2.71$\\
$-5\times10^{-5}$       & $2.171$ & $4.104$ & $89.06$  & $86.81$  & $2.25$\\
$-10^{-4}$             & $3.619$ & $6.262$ & $73.02$  & $71.15$  & $1.86$\\
$-2\times10^{-4}$      & $10.45$ & $15.36$ & $46.91$  & $45.66$  & $1.25$\\
$-3\times10^{-4}$      & $32.78$ & $41.86$ & $27.72$  & $26.89$  & $0.83$\\
$-4\times10^{-4}$      & $122.2$ & $139.6$ & $14.23$  & $13.63$  & $0.59$\\
\end{tabular}
\end{ruledtabular}
\end{table}

\begin{figure}[!htbp]
\centering
\includegraphics[width=0.99\textwidth]{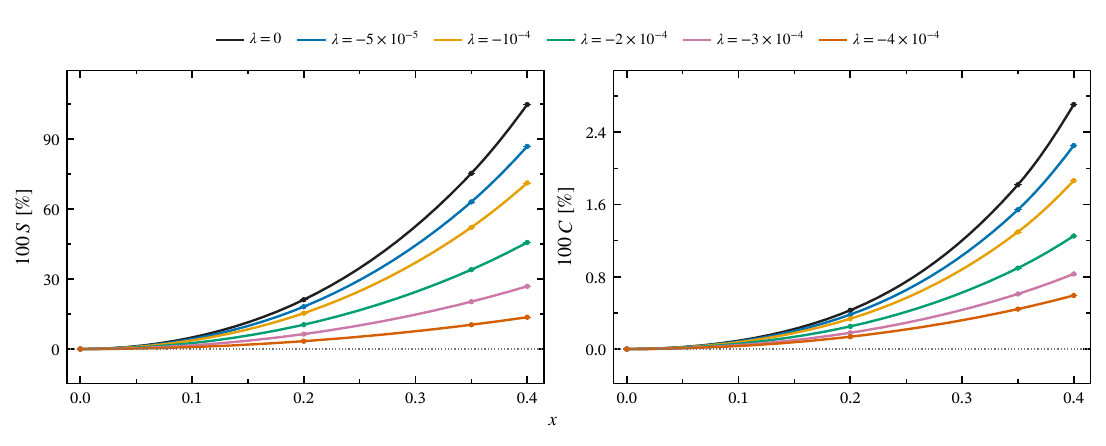}
\caption{Decomposition of the tidal change at fixed
$\log_{10}A=57.20715121304262$ into structural $100S$ (left)
and explicit mixed $100C$ (right) contributions defined in
Eq.~\eqref{eq:results-tidal-decomposition}. Colors identify
the six gravitational couplings. Both components use the
static deformability at the same $A$ and coupling, with
$\delta_{\rm rot}\lambda_{\rm T}=S+C$.
Markers and bars give the four calculated spins and their
separate empirical control envelopes. PCHIP interpolation
in $x^2$ is applied to the total and $C$, with displayed $S$
obtained by subtraction to preserve the sum between nodes.
The panels have different vertical scales; interpolated
points are not additional equilibrium models.}
\label{fig:tidal-decomposition}
\end{figure}

The relation to the density redistribution becomes explicit
in Fig.~\ref{fig:tidal-decomposition}. Let
$\lambda_{\rm ref}=\lambda_{\rm T}(A,0)$ and denote by
$\lambda_{\rm static}(p_{c,\Omega})$ the static deformability
of the spherical background associated with a rotating
equilibrium. The retained response is the sum of this term
and the explicit mixed correction
$\delta\lambda_{\rm mixed}(p_{c,\Omega},\Omega)$. The normalized
decomposition of Eq.~\eqref{eq:th-decomposition} then reads
\begin{subequations}\label{eq:results-tidal-decomposition}
\begin{align}
 S&=\frac{\lambda_{\rm static}(p_{c,\Omega})}
          {\lambda_{\rm ref}}-1,\\
C&=\frac{\delta\lambda_{\rm mixed}(p_{c,\Omega},\Omega)}
          {\lambda_{\rm ref}},\\
\delta_{\rm rot}\lambda_{\rm T}&=S+C .
\end{align}
\end{subequations}
The spherical background at $p_{c,\Omega}$ is not the initial
static star at fixed $A$. Its changing density and radius
account for $S$, while $C$ is the remaining explicit
rotational tidal correction in the adopted perturbative
prescription. The components depend on that prescription;
they are not independently defined observables.

Most of the fixed-$A$ increase comes from the structural
term. At $x=0.40$, Table~\ref{tab:tidal-fixed-A} shows that $S$ supplies approximately
$95.8$--$97.5\%$ of the total increase across the six
couplings. In GR the contributions are $100S=104.73$ and
$100C=2.71$, whereas at $\lambda=-4\times10^{-4}$ they are
$13.63$ and $0.59$, respectively. These are contributions
in percentage points to the total enhancement, not
percentages of a radius change. The positive mixed term
is resolved relative to its empirical control envelope at
the nonzero calculated spins, even though it contributes
a smaller fraction of the total. Its inclusion is required
to retain the selected response consistently through
second order in rotation.

The structural identity in Eq.~\eqref{eq:th-structural-identity}
quantifies how the radius and Love number changes contribute to $S$.
For the GR sequence at $x=0.40$, the background radius
increases by about $13.2\%$ and its $k_2$ by $10.3\%$.
The radius factor alone is approximately $1.856$; multiplying
it by the $k_2$ ratio gives $1+S\simeq2.047$.
At $\lambda=-4\times10^{-4}$, the corresponding radius
increase is only about $3.54\%$ and the background $k_2$
decreases by $4.50\%$, yielding $1+S\simeq1.136$.
The smaller fractional tidal enhancement at this coupling
is therefore accompanied by a smaller fractional expansion
of the associated spherical background and a different
change in its dimensionless Love number. This identity
quantifies the structural explanation without assuming
that the Love number remains constant or replacing the
tidal calculation by a fitted power law.

The dominance of the structural contribution has an important implication for interpreting tidal deformability as a probe of gravity. Since the dimensional response scales with the fifth power of the stellar radius, a substantial departure from the GR prediction can arise primarily from changes in equilibrium size, without a comparably large change in the dimensionless Love number. The coupling dependence of the deformability should therefore be interpreted together with the corresponding mass, radius and rotational state. In this respect, the Love number provides complementary information by removing the explicit fifth-power radius factor, although it remains sensitive to the stellar density profile. Establishing how much independent information the tidal response carries about the gravitational coupling consequently requires separating these structural effects before assessing observational distinguishability.

The present model also illustrates how stellar tides can probe gravitational modifications that act through matter rather than through additional vacuum degrees of freedom. For $f(R,T)=R+2\lambda T$, the exterior vacuum equations coincide with those of GR, while the coupling to the matter trace changes the interior equilibrium and perturbation equations. Its influence on the tidal response is therefore encoded in the stellar configuration and in the matching of the interior solution to the exterior field. A GR vacuum exterior consequently does not imply a GR tidal deformability. However, because the interior equations admit an effective-fluid representation, attributing a measured departure specifically to the gravitational coupling would require independent constraints on the physical matter model.

Normalization also matters for the explicit mixed term.
Although $C$ decreases toward more negative coupling,
$\delta\lambda_{\rm mixed}$ at $x=0.40$ increases from
approximately $3.56\times10^{13}\ \mathrm{km}^5$ in GR
to $7.25\times10^{14}\ \mathrm{km}^5$ at
$\lambda=-4\times10^{-4}$. A smaller $C$ consequently
does not demonstrate suppression of the absolute mixed
response. The large total percentages likewise describe
changes in a response coefficient; they are not the
geometrical amplitude of an imposed tide. The latter
also depends on the external field strength, which
does not enter the fractional comparison in
Eq.~\eqref{eq:results-fixed-A-change}.

\begin{figure}[!htbp]
\centering
\includegraphics[width=0.99\textwidth]{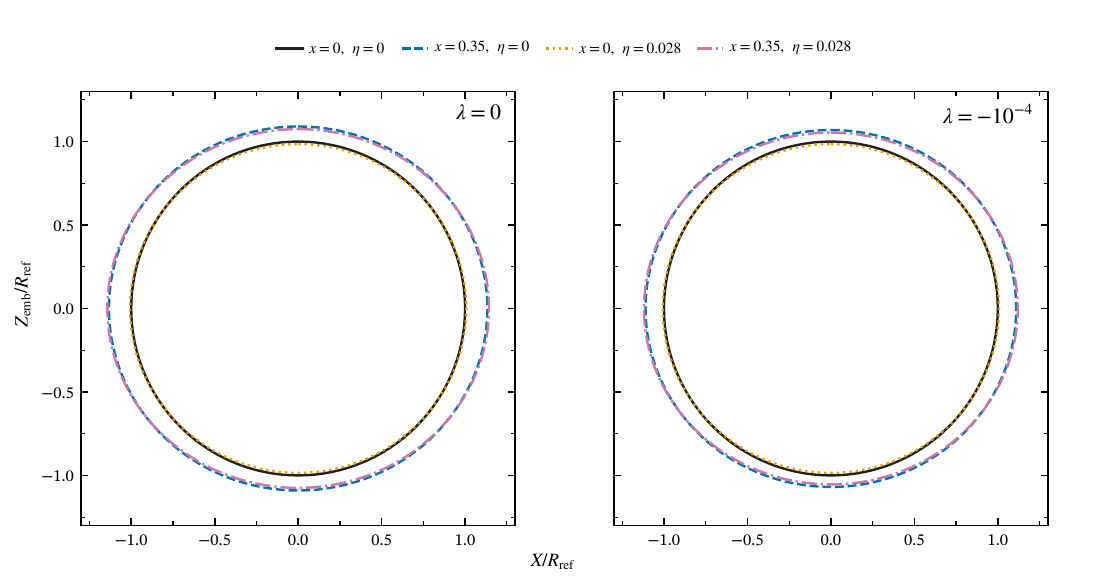}
\caption{Meridional embeddings of the spatial two-metric on the
operational material isobar at fixed
$\log_{10}A=57.20715121304262$. Panels give $\lambda=0$ and $-10^{-4}$.
$X$ is perpendicular and $Z_{\rm emb}$ parallel to the spin axis; both are
normalized by the panel's unperturbed nonrotating radius $R_{\rm ref}$.
The legend identifies $(x,\eta)=(0,0),(0.35,0),(0,0.028),(0.35,0.028)$,
where $x$ labels the isolated background and $\eta$ is the prescribed
positive field strength in Eq.~\eqref{eq:results-surface-forcing}.
The value $0.028$ follows from the equal-mass periastron scale in
Ref.~\cite{McNeillMardlingMueller2020}.
$A,J$ are conserved during the tidal perturbation. The contours include
the displaced surface and angular metric through
$\Omega^2,\mathcal E,\mathcal E\Omega^2$, without deformation amplification.
The embedded polar height is distinct from a coordinate polar radius.
These surfaces represent the selected stationary $m=0$ tide, with no
elastic response or additional binary components.}
\label{fig:tidal-surface-geometry}
\end{figure}

The distinction between a response coefficient and a finite surface
distortion can be made explicit using the same displaced isobar as in
the matching calculation and the embedding in
Eq.~\eqref{eq:th-surface-embedding}. Figure~\ref{fig:tidal-surface-geometry}
compares the isolated and tidally perturbed meridional surfaces at
$\lambda=0$ and $-10^{-4}$, with the same baryon target as
Figs.~\ref{fig:density-redistribution}, \ref{fig:tidal-fixed-A}, and
\ref{fig:tidal-decomposition}. The applied field is specified by
\begin{equation}
 \eta=\frac{\mathcal E R_0^3}{M_0}
 \simeq\frac{M_{\rm comp}}{M_0}\left(\frac{R_0}{d_{\rm orb}}\right)^3,
 \label{eq:results-surface-forcing}
\end{equation}
where $M_0,R_0$ refer to the associated spherical background,
$M_{\rm comp}$ is a companion mass, and $d_{\rm orb}$ is the separation in
the leading Newtonian external-field interpretation. The last relation
is the stationary $m=0$ projection of the quadrupolar companion
potential with the normalization used here \cite{YuWeinbergFuller2020}.
We prescribe $\eta=0.028$, rounded from $1/(3.3)^3$ using the equal-mass,
periastron-separation example in Fig.~1 of
McNeill et al.~\cite{McNeillMardlingMueller2020}.
Only this dimensionless forcing scale is adopted from that theoretical
configuration; the masses, radii, and orbital evolution of that binary
are not assigned to our sequence. Equal $\eta$ also does not imply
equal dimensional $\mathcal E$ across the stellar backgrounds.

For the positive field used here, the leading quadrupolar acceleration
is outward in the equatorial plane and inward along the spin axis.
The tidal perturbation therefore adds flattening to the isolated
rotational distortion. At $x=0.35$, switching from $\eta=0$ to $0.028$
increases the equatorial circumferential radius by approximately
$0.84\%$ in GR and $0.83\%$ at $\lambda=-10^{-4}$, while the embedded
polar height decreases by $1.44\%$ and $1.43\%$, respectively.
These changes compare surfaces at the same background spin; they
should not be confused with the larger expansion between the initial
static star and the rotating equilibrium. The polar height of each
rotating, tidally perturbed model remains above its own $R_{\rm ref}$,
consistent with the overall expansion already illustrated by
Fig.~\ref{fig:density-redistribution}.

The two normalized panels have similar shapes even though their
reference radii are approximately $2172$ and $2762\ \mathrm{km}$.
Removing this dimensional scale from the axes does not remove its
fifth-power contribution to the static deformability, quantified in
Eq.~\eqref{eq:th-structural-identity}. The surface comparison thus
connects the density rearrangement to the tidal analysis while keeping
their physical roles distinct: Fig.~\ref{fig:density-redistribution}
compares isolated density profiles, Fig.~\ref{fig:tidal-decomposition}
separates changes in the induced-multipole response, and
Fig.~\ref{fig:tidal-surface-geometry} displays the surface deformation
for a specified applied field. A large fractional change of
$\lambda_{\rm T}$ can consequently coexist with a small additional
geometric distortion at the chosen forcing. The contours visualize the
retained surface geometry; the tidal coefficient continues to be
extracted from the induced exterior multipole, rather than fitted from
the outline.

Only the four marked spins in Figs.~\ref{fig:tidal-fixed-A}
and \ref{fig:tidal-decomposition}
represent calculated configurations. The interpolation
in $x^2$ provides an even visual continuation under spin
reversal, preserves the calculated values, and maintains
the component sum. It neither supplies new stellar
solutions nor constrains the omitted fourth-order
rotational terms. For the tidal population shown here,
the largest admitted $\epsilon_{\rm rot}$ is about
$0.0997$; the empirical envelopes quantify numerical,
matching, surface, and baryon-pairing sensitivities
within the retained model. The surface embedding uses the same
perturbative orders; its geometric consistency checks do not supply
bounds on omitted $\mathcal E^2$ or $\Omega^4$ terms.

All of these tidal comparisons describe cold fluid
matter, including its lattice contribution to the EOS
but no elastic shear response. Perot and
Chamel~\cite{PerotChamel2022} found that elasticity
reduces the tidal deformability of crystallized white
dwarfs, particularly at low masses. Their result
identifies a distinct material contribution that is
absent from the present fluid calculation; agreement
with a crystallized model should therefore not be
assumed from a common nuclear composition alone.
The coupling dependence and structural dominance
established here apply to the selected EOS, surface
prescription, and baryon sequence. Extending the
comparison to other values of $A$ would test how
widely this separation persists. Relating it to an
observed binary would additionally require the relevant
tidal components and orbital dynamics, beyond the
stationary aligned response calculated in this work.

\section{Conclusions}\label{sec:conclusion}

We have calculated the structure and stationary electric tidal response
of cold $^{12}\mathrm C$ fluid white dwarfs in
$f(R,T)=R+2\lambda T$ gravity, retaining uniform rotation through second
order. Varying the gravitational coupling while preserving the material
EOS and surface prescription separates the change in gravitational
equilibrium from changes in composition. Negative coupling produces
larger radii over the common mass interval and increases the dimensional
tidal deformability. The effective pressure balance explains why even
small couplings can affect the extended stellar layers without a change
in the microscopic EOS. Rotation adds a positive mass correction at
prescribed central density, expands the equatorial surface, and generates
an isolated mass quadrupole. These mass, shape, and multipole corrections
have distinct dependences on the reference mass and coupling.

Identification by physical baryon number is necessary to interpret how
the response changes when the amount of matter is conserved. In the
representative sequence at $\lambda=-10^{-4}$, increasing rotational
support lowers the central density and expands both the equatorial and
polar radii relative to the initial static star, while the configuration
remains oblate. This adjustment also changes the background on which the
tide acts. Along the common baryon target studied here, the dimensional
deformability increases at successive calculated spins for all six
couplings. At $x=\Omega/\Omega_K=0.40$, its enhancement relative to each
coupling's own static configuration is approximately $107.4\%$ in GR and
$14.2\%$ at $\lambda=-4\times10^{-4}$. A smaller fractional enhancement
at negative coupling therefore coexists with a larger absolute tidal
response than in GR.

Changes in the associated spherical background supply
$95.8$--$97.5\%$ of the enhancement at $x=0.40$ in the adopted
prescription. The fifth-power radius dependence and the changing
background Love number together account for this structural contribution;
holding the Love number constant would miss part of the adjustment.
The explicit mixed rotational tidal term is positive and resolved by
the numerical controls, although it contributes a smaller fraction of
the total. At this spin, its dimensional value increases toward more
negative coupling on this sequence, even as its normalized contribution
decreases.
The separation thus distinguishes a change in stellar structure from
the remaining explicit coupling between rotation and the applied tide.
It is conditional on the background and response conventions and does
not define two separately measurable tidal observables.

Surface embeddings give a complementary interpretation of the induced
multipole response. At $x=0.35$, imposing the prescribed field
$\eta=0.028$ produces additional surface changes of order one percent
for both GR and $\lambda=-10^{-4}$. Large fractional changes in a
deformability coefficient can therefore accompany a small additional
distortion at a specified forcing strength. The similar normalized
contours at the two couplings also coexist with different dimensional
responses, since removing the reference radius from the plotted
coordinates does not remove its contribution to the deformability.
The displaced material surface and its angular metric are needed to
reconstruct that geometry; the induced quadrupole alone does not fix
the outline.

Quantitative tidal conclusions apply to the admitted models with
$x\leq0.40$ and $\epsilon_{\rm rot}\leq0.1$. The structural curves
continued to the Kepler reference illustrate the retained expansion,
without establishing its accuracy near mass shedding. Comparison with
calculations that retain the full rotational deformation would test
how far the structural and tidal trends persist at higher spin. Within
the controlled domain, repeating the decomposition at several baryon
targets would test whether the structural dominance depends on the
stellar mass or is shared by a wider set of equilibria. Extending the
material description beyond cold carbon, and including elastic shear
where crystallization is relevant, would assess how composition and
material stresses compete with the coupling dependence found here.
These comparisons would also distinguish changes in the dimensionless
internal response from those associated with the stellar radius and
the adopted material surface.

A dynamical extension must address a different part of the tidal
problem. The present calculation determines the stationary aligned
electric $\ell=2,m=0$ response at fixed $A,J$; it does not calculate
oscillation frequencies, damping, or resonant excitation. Constructing
time-dependent fluid perturbations consistent with the chosen matter
and gravity prescriptions, and determining the frequency dependence of
the relevant azimuthal components, would test whether the coupling
dependence of the equilibrium response persists near stellar resonances.
Combining that calculation with orbital evolution would permit a
comparison with binary signals. Assessing whether such signals could
distinguish $\lambda$ from changes in mass, spin, and composition would
then require their joint uncertainties. Such a comparison would test
whether the mixed response adds information beyond the structural
changes identified in the present equilibrium sequences.

\begin{acknowledgments}
Edson Otoniel acknowledges support from Funda\c{c}\~ao Cearense de Apoio ao Desenvolvimento Cient\'ifico e Tecnol\'ogico (FUNCAP) through grant BP6-0241-00335.01.00/25. PHRSM would like to thank CNPq (Conselho Nacional de Desenvolvimento
Científico e Tecnológico) for partial financial support under grant No.
310366/2023-2.
\end{acknowledgments}

%\appendix
%\section{Perturbation equations and matching conventions}\label{app:perturbations}
% Slot for long equations, unit/sign conventions and interface/surface terms.

%\section{Numerical and surface-sensitivity controls}\label{app:controls}
% Slot for benchmarks, actual settings, pointwise envelopes and branch diagnostics.

% Earlier static dimensionless figure omitted; original assets preserved.

% Verified journal references cited in Results and Discussion.
\bibliographystyle{apsrev4-2}
\bibliography{ref}

@ARTICLE{althaus/2010,
       author = {{Althaus}, Leandro G. and {C{\'o}rsico}, Alejandro H. and
                 {Isern}, Jordi and {Garc{\'i}a-Berro}, Enrique},
        title = "{Evolutionary and pulsational properties of white dwarf stars}",
      journal = {\aapr},
         year = 2010,
        month = oct,
       volume = {18},
       number = {4},
        pages = {471-566},
          doi = {10.1007/s00159-010-0033-1},
archivePrefix = {arXiv},
       eprint = {1007.2659},
 primaryClass = {astro-ph.SR},
       adsurl = {https://ui.adsabs.harvard.edu/abs/2010A&ARv..18..471A}
}

@ARTICLE{maoz/2014,
       author = {{Maoz}, Dan and {Mannucci}, Filippo and {Nelemans}, Gijs},
        title = "{Observational Clues to the Progenitors of Type Ia Supernovae}",
      journal = {\araa},
         year = 2014,
       volume = {52},
        pages = {107-170},
          doi = {10.1146/annurev-astro-082812-141031},
archivePrefix = {arXiv},
       eprint = {1312.0628},
 primaryClass = {astro-ph.CO},
       adsurl = {https://ui.adsabs.harvard.edu/abs/2014ARA&A..52..107M}
}

@ARTICLE{riess/1998,
       author = {{Riess}, Adam G. and others},
        title = "{Observational Evidence from Supernovae for an Accelerating Universe and a Cosmological Constant}",
      journal = {\aj},
         year = 1998,
       volume = {116},
       number = {3},
        pages = {1009-1038},
          doi = {10.1086/300499},
archivePrefix = {arXiv},
       eprint = {astro-ph/9805201},
       adsurl = {https://ui.adsabs.harvard.edu/abs/1998AJ....116.1009R}
}

@ARTICLE{perlmutter/1999,
       author = {{Perlmutter}, S. and others},
        title = "{Measurements of {$\Omega$} and {$\Lambda$} from 42 High-Redshift Supernovae}",
      journal = {\apj},
         year = 1999,
       volume = {517},
       number = {2},
        pages = {565-586},
          doi = {10.1086/307221},
archivePrefix = {arXiv},
       eprint = {astro-ph/9812133},
       adsurl = {https://ui.adsabs.harvard.edu/abs/1999ApJ...517..565P}
}

@ARTICLE{korol/2017,
       author = {{Korol}, Valeriya and {Rossi}, Elena M. and
                 {Groot}, Paul J. and {Nelemans}, Gijs and
                 {Toonen}, Silvia and {Brown}, Anthony G. A.},
        title = "{Prospects for detection of detached double white dwarf binaries with Gaia, LSST and LISA}",
      journal = {\mnras},
         year = 2017,
        month = sep,
       volume = {470},
       number = {2},
        pages = {1894-1910},
          doi = {10.1093/mnras/stx1285},
archivePrefix = {arXiv},
       eprint = {1703.02555},
 primaryClass = {astro-ph.HE},
       adsurl = {https://ui.adsabs.harvard.edu/abs/2017MNRAS.470.1894K}
}

@ARTICLE{kupfer/2024,
       author = {{Kupfer}, Thomas and {Korol}, Valeriya and
                 {Littenberg}, Tyson B. and {Shah}, Sweta and
                 {Savalle}, Etienne and {Groot}, Paul J. and
                 {Marsh}, Thomas R. and {Le Jeune}, Maude and
                 {Nelemans}, Gijs and {Pala}, Anna F. and
                 {Petiteau}, Antoine and {Ramsay}, Gavin and
                 {Steeghs}, Danny and {Babak}, Stanislav},
        title = "{LISA Galactic Binaries with Astrometry from Gaia DR3}",
      journal = {\apj},
         year = 2024,
        month = mar,
       volume = {963},
       number = {2},
          eid = {100},
        pages = {100},
          doi = {10.3847/1538-4357/ad2068},
archivePrefix = {arXiv},
       eprint = {2302.12719},
 primaryClass = {astro-ph.SR},
       adsurl = {https://ui.adsabs.harvard.edu/abs/2024ApJ...963..100K}
}

@article{Carvalho2017,
  author = {Carvalho, G. A. and Lobato, R. V. and Moraes, P. H. R. S.
            and Arba{\~n}il, Jos{\'e} D. V. and Otoniel, E.
            and Marinho, Jr., R. M. and Malheiro, M.},
  title = {Stellar equilibrium configurations of white dwarfs in the {$f(R,T)$} gravity},
  journal = {Eur. Phys. J. C},
  volume = {77},
  number = {12},
  pages = {871},
  year = {2017},
  doi = {10.1140/epjc/s10052-017-5413-5}
}

@ARTICLE{otoniel/2026,
       author = {{Otoniel}, Edson and {Pretel}, Juan M.~Z. and {Mota}, Cl{\'e}sio E. and {Flores}, C{\'e}sar O.~V. and {Alves}, Victor B.~T. and {da Silva}, Franciele M.},
        title = "{White dwarf structure in f(R, T, L$_{m}$) gravity: Beyond the Chandrasekhar mass limit}",
      journal = {Physics Letters B},
         year = 2026,
        month = apr,
       volume = {875},
          eid = {140323},
        pages = {140323},
          doi = {10.1016/j.physletb.2026.140323},
archivePrefix = {arXiv},
       eprint = {2507.18745},
 primaryClass = {gr-qc},
       adsurl = {https://ui.adsabs.harvard.edu/abs/2026PhLB..87540323O}
}

@ARTICLE{kalita/2021,
       author = {{Kalita}, Surajit and {Mukhopadhyay}, Banibrata},
        title = "{Gravitational Wave in f(R) Gravity: Possible Signature of Sub- and Super-Chandrasekhar Limiting-mass White Dwarfs}",
      journal = {\apj},
         year = 2021,
        month = mar,
       volume = {909},
       number = {1},
          eid = {65},
        pages = {65},
          doi = {10.3847/1538-4357/abddb8},
archivePrefix = {arXiv},
       eprint = {2101.07278},
 primaryClass = {astro-ph.HE},
       adsurl = {https://ui.adsabs.harvard.edu/abs/2021ApJ...909...65K}
}

@ARTICLE{li/2024,
       author = {{Li}, Jie and {Yang}, Bo and {Lin}, Wenbin},
        title = "{Massive white dwarfs in Rastall-Rainbow gravity}",
      journal = {\jcap},
         year = 2024,
        month = apr,
       volume = {2024},
       number = {4},
          eid = {081},
        pages = {081},
          doi = {10.1088/1475-7516/2024/04/081},
archivePrefix = {arXiv},
       eprint = {2305.17676},
 primaryClass = {gr-qc},
       adsurl = {https://ui.adsabs.harvard.edu/abs/2024JCAP...04..081L}
}

@ARTICLE{das/2015,
       author = {{Das}, Upasana and {Mukhopadhyay}, Banibrata},
        title = "{Modified Einstein's gravity as a possible missing link between sub- and super-Chandrasekhar type Ia supernovae}",
      journal = {\jcap},
         year = 2015,
        month = may,
       volume = {2015},
       number = {5},
        pages = {045-045},
          doi = {10.1088/1475-7516/2015/05/045},
archivePrefix = {arXiv},
       eprint = {1411.1515},
 primaryClass = {astro-ph.SR},
       adsurl = {https://ui.adsabs.harvard.edu/abs/2015JCAP...05..045D}
}

@ARTICLE{taylor/2020,
       author = {{Taylor}, Andrew J. and {Yagi}, Kent and {Arras}, Phil},
        title = "{I-Love-Q relations for realistic white dwarfs}",
      journal = {\mnras},
         year = 2020,
        month = feb,
       volume = {492},
       number = {1},
        pages = {978-992},
          doi = {10.1093/mnras/stz3519},
archivePrefix = {arXiv},
       eprint = {1912.09557},
 primaryClass = {gr-qc},
       adsurl = {https://ui.adsabs.harvard.edu/abs/2020MNRAS.492..978T}
}

@ARTICLE{chandrasekhar/1931,
       author = {{Chandrasekhar}, S.},
        title = "{The Maximum Mass of Ideal White Dwarfs}",
      journal = {\apj},
         year = 1931,
        month = jul,
       volume = {74},
        pages = {81-82},
          doi = {10.1086/143324},
       adsurl = {https://ui.adsabs.harvard.edu/abs/1931ApJ....74...81C}
}

@ARTICLE{howell/2006,
       author = {{Howell}, D. Andrew and others},
        title = "{The type Ia supernova SNLS-03D3bb from a super-Chandrasekhar-mass white dwarf star}",
      journal = {\nat},
         year = 2006,
        month = sep,
       volume = {443},
       number = {7109},
        pages = {308-311},
          doi = {10.1038/nature05103},
archivePrefix = {arXiv},
       eprint = {astro-ph/0609616},
       adsurl = {https://ui.adsabs.harvard.edu/abs/2006Natur.443..308H}
}

@ARTICLE{scalzo/2010,
       author = {{Scalzo}, R. A. and others},
        title = "{Nearby Supernova Factory Observations of SN 2007if: First Total Mass Measurement of a Super-Chandrasekhar-Mass Progenitor}",
      journal = {\apj},
         year = 2010,
        month = apr,
       volume = {713},
       number = {2},
        pages = {1073-1094},
          doi = {10.1088/0004-637X/713/2/1073},
archivePrefix = {arXiv},
       eprint = {1003.2217},
 primaryClass = {astro-ph.CO},
       adsurl = {https://ui.adsabs.harvard.edu/abs/2010ApJ...713.1073S}
}

@ARTICLE{taubenberger/2011,
       author = {{Taubenberger}, S. and others},
        title = "{High luminosity, slow ejecta and persistent carbon lines: SN 2009dc challenges thermonuclear explosion scenarios}",
      journal = {\mnras},
         year = 2011,
        month = apr,
       volume = {412},
       number = {4},
        pages = {2735-2762},
          doi = {10.1111/j.1365-2966.2010.18107.x},
archivePrefix = {arXiv},
       eprint = {1011.5665},
 primaryClass = {astro-ph.SR},
       adsurl = {https://ui.adsabs.harvard.edu/abs/2011MNRAS.412.2735T}
}

@ARTICLE{moraes/2016,
       author = {{Moraes}, P.~H.~R.~S. and {Ribeiro}, G. and {Correa}, R.~A.~C.},
        title = "{A transition from a decelerated to an accelerated phase of the universe expansion from the simplest non-trivial polynomial function of T in the f(R,T) formalism}",
      journal = {\apss},
         year = 2016,
        month = jul,
       volume = {361},
       number = {7},
          eid = {227},
        pages = {227},
          doi = {10.1007/s10509-016-2788-y},
archivePrefix = {arXiv},
       eprint = {1602.07159},
 primaryClass = {hep-th},
       adsurl = {https://ui.adsabs.harvard.edu/abs/2016Ap&SS.361..227M}
}

@ARTICLE{sahoo/2018,
       author = {{Sahoo}, P.~K. and {Moraes}, P.~H.~R.~S. and {Sahoo}, Parbati and {Bishi}, Binaya K.},
        title = "{f(R,T)=f(R)+{\ensuremath{\lambda}}T gravity models as alternatives to cosmic acceleration}",
      journal = {European Physical Journal C},
         year = 2018,
        month = sep,
       volume = {78},
       number = {9},
          eid = {736},
        pages = {736},
          doi = {10.1140/epjc/s10052-018-6211-4},
archivePrefix = {arXiv},
       eprint = {1809.03303},
 primaryClass = {gr-qc},
       adsurl = {https://ui.adsabs.harvard.edu/abs/2018EPJC...78..736S}
}

@ARTICLE{myrzakulov/2025,
       author = {{Myrzakulov}, N. and {Shekh}, S.~H. and {Pradhan}, Anirudh and {Dixit}, Archana},
        title = "{Dark energy and cosmic evolution: A study in f(R,T) gravity}",
      journal = {Journal of High Energy Astrophysics},
         year = 2025,
        month = jul,
       volume = {47},
          eid = {100374},
        pages = {100374},
          doi = {10.1016/j.jheap.2025.100374},
archivePrefix = {arXiv},
       eprint = {2501.08362},
 primaryClass = {gr-qc},
       adsurl = {https://ui.adsabs.harvard.edu/abs/2025JHEAp..4700374M}
}

@ARTICLE{sharif/2018,
       author = {{Sharif}, M. and {Siddiqa}, Aisha},
        title = "{Study of stellar structures in f(R,T) gravity}",
      journal = {International Journal of Modern Physics D},
         year = 2018,
        month = jan,
       volume = {27},
       number = {7},
          eid = {1850065},
        pages = {1850065},
          doi = {10.1142/S0218271818500657},
       adsurl = {https://ui.adsabs.harvard.edu/abs/2018IJMPD..2750065S}
}

@ARTICLE{chowdhury/2025,
       author = {{Chowdhury}, Anupama Roy and {Das}, Shyam and {Rahaman}, Farook},
        title = "{A Study of Compact Stellar Objects in f(R, T) Theory of Gravity}",
      journal = {Universe},
         year = 2025,
        month = dec,
       volume = {11},
       number = {12},
          eid = {409},
        pages = {409},
          doi = {10.3390/universe11120409},
       adsurl = {https://ui.adsabs.harvard.edu/abs/2025Univ...11..409C}
}

@ARTICLE{carvalho/2020,
       author = {{Carvalho}, G.~A. and {Dos Santos}, S.~I. and {Moraes}, P.~H.~R.~S. and {Malheiro}, M.},
        title = "{Strange stars in energy-momentum-conserved f(R,T) gravity}",
      journal = {International Journal of Modern Physics D},
         year = 2020,
        month = jan,
       volume = {29},
       number = {10},
          eid = {2050075},
        pages = {2050075},
          doi = {10.1142/S0218271820500753},
archivePrefix = {arXiv},
       eprint = {1911.02484},
 primaryClass = {gr-qc},
       adsurl = {https://ui.adsabs.harvard.edu/abs/2020IJMPD..2950075C}
}

@ARTICLE{deb/2019,
       author = {{Deb}, Debabrata and {Ketov}, Sergei V. and {Maurya}, S.~K. and {Khlopov}, Maxim and {Moraes}, P.~H.~R.~S. and {Ray}, Saibal},
        title = "{Exploring physical features of anisotropic strange stars beyond standard maximum mass limit in f(R,T) gravity}",
      journal = {\mnras},
         year = 2019,
        month = jun,
       volume = {485},
       number = {4},
        pages = {5652-5665},
          doi = {10.1093/mnras/stz708},
archivePrefix = {arXiv},
       eprint = {1810.07678},
 primaryClass = {gr-qc},
       adsurl = {https://ui.adsabs.harvard.edu/abs/2019MNRAS.485.5652D}
}

@ARTICLE{dos_santos/2019,
       author = {{dos Santos}, S.~I. and {Carvalho}, G.~A. and {Moraes}, P.~H.~R.~S. and {Lenzi}, C.~H. and {Malheiro}, M.},
        title = "{A conservative energy-momentum tensor in the f(R,T) gravity and its implications for the phenomenology of neutron stars}",
      journal = {European Physical Journal Plus},
         year = 2019,
        month = aug,
       volume = {134},
       number = {8},
          eid = {398},
        pages = {398},
          doi = {10.1140/epjp/i2019-12830-8},
archivePrefix = {arXiv},
       eprint = {1803.07719},
 primaryClass = {gr-qc},
       adsurl = {https://ui.adsabs.harvard.edu/abs/2019EPJP..134..398D}
}

@ARTICLE{feng/2025,
       author = {{Feng}, Zhe},
        title = "{Charged anisotropic white dwarfs in f(R,T) gravity}",
      journal = {Modern Physics Letters A},
         year = 2025,
        month = feb,
       volume = {40},
       number = {4},
          eid = {2450218},
        pages = {2450218},
          doi = {10.1142/S0217732324502183},
archivePrefix = {arXiv},
       eprint = {2210.01574},
 primaryClass = {gr-qc},
       adsurl = {https://ui.adsabs.harvard.edu/abs/2025MPLA...4050218F}
}

@ARTICLE{rocha/2020,
       author = {{Rocha}, F. and {Carvalho}, G.~A. and {Deb}, D. and {Malheiro}, M.},
        title = "{Study of the charged super-Chandrasekhar limiting mass white dwarfs in the f (R ,T ) gravity}",
      journal = {\prd},
         year = 2020,
        month = may,
       volume = {101},
       number = {10},
          eid = {104008},
        pages = {104008},
          doi = {10.1103/PhysRevD.101.104008},
archivePrefix = {arXiv},
       eprint = {1911.08894},
 primaryClass = {physics.gen-ph},
       adsurl = {https://ui.adsabs.harvard.edu/abs/2020PhRvD.101j4008R}
}

@ARTICLE{bhattacharjee/2024,
       author = {{Bhattacharjee}, Snehasish},
        title = "{White dwarf cooling in f(R,T) gravity}",
      journal = {International Journal of Modern Physics A},
         year = 2024,
        month = feb,
       volume = {39},
          eid = {2450026},
        pages = {2450026},
          doi = {10.1142/S0217751X2450026X},
archivePrefix = {arXiv},
       eprint = {2305.10144},
 primaryClass = {gr-qc},
       adsurl = {https://ui.adsabs.harvard.edu/abs/2024IJMPA..3950026B}
}

@article{Otoniel2021Rotation,
  author = {Otoniel, Edson and Coelho, Jaziel G. and Nunes, S{\'i}lvia P. and Malheiro, Manuel and Weber, Fridolin},
  title = {Mass limits of the extremely fast-spinning white dwarf {CTCV J2056--3014}},
  journal = {Astronomy \& Astrophysics},
  volume = {656},
  pages = {A77},
  year = {2021},
  doi = {10.1051/0004-6361/202039749}
}

@article{Otoniel2026StrangeRotation,
  author = {da Silva, Edson Otoniel and Arba{\~n}il, Jos{\'e} D. V. and Carvalho, Geanderson A. and Weber, Fridolin},
  title = {Rotating strange dwarfs and their indistinguishability from white dwarfs},
  journal = {The European Physical Journal C},
  volume = {86},
  pages = {393},
  year = {2026},
  doi = {10.1140/epjc/s10052-026-15666-4}
}

@article{BinningtonPoisson2009,
  author = {Binnington, Taylor and Poisson, Eric},
  title = {Relativistic theory of tidal {Love} numbers},
  journal = {Phys. Rev. D},
  volume = {80},
  number = {8},
  pages = {084018},
  year = {2009},
  doi = {10.1103/PhysRevD.80.084018}
}

@article{Pani2015,
  author = {Pani, Paolo and Gualtieri, Leonardo and Maselli, Andrea and Ferrari, Valeria},
  title = {Tidal deformations of a spinning compact object},
  journal = {Phys. Rev. D},
  volume = {92},
  number = {2},
  pages = {024010},
  year = {2015},
  doi = {10.1103/PhysRevD.92.024010}
}

@article{MurshidKalam2024,
  author = {Murshid, Masum and Kalam, Mehedi},
  title = {Neutron stars in {$f(R,T)$} theory: slow rotation approximation},
  journal = {J. Cosmol. Astropart. Phys.},
  volume = {2024},
  number = {09},
  pages = {030},
  year = {2024},
  doi = {10.1088/1475-7516/2024/09/030}
}

@article{PerotChamel2022,
  author = {Perot, L. and Chamel, N.},
  title = {Tidal deformability of crystallized white dwarfs in full general relativity},
  journal = {Phys. Rev. D},
  volume = {106},
  number = {2},
  pages = {023012},
  year = {2022},
  doi = {10.1103/PhysRevD.106.023012}
}

@article{Harko2011,
  author = {Harko, Tiberiu and Lobo, Francisco S. N. and Nojiri, Shin'ichi
            and Odintsov, Sergei D.},
  title = {{$f(R,T)$} gravity},
  journal = {Phys. Rev. D},
  volume = {84},
  pages = {024020},
  year = {2011},
  doi = {10.1103/PhysRevD.84.024020}
}

@article{Chandrasekhar1931,
  author = {Chandrasekhar, S.},
  title = {The Maximum Mass of Ideal White Dwarfs},
  journal = {Astrophys. J.},
  volume = {74},
  pages = {81--82},
  year = {1931},
  doi = {10.1086/143324}
}

@article{Salpeter1961,
  author = {Salpeter, E. E.},
  title = {Energy and Pressure of a Zero-Temperature Plasma},
  journal = {Astrophys. J.},
  volume = {134},
  pages = {669},
  year = {1961},
  doi = {10.1086/147194}
}

@article{ChamelFantina2015Capture,
  author = {Chamel, N. and Fantina, A. F.},
  title = {Electron capture instability in magnetic and nonmagnetic white dwarfs},
  journal = {Phys. Rev. D},
  volume = {92},
  pages = {023008},
  year = {2015},
  doi = {10.1103/PhysRevD.92.023008}
}

@article{Chamel2015NeutronDrip,
  author = {Chamel, N. and Fantina, A. F. and Zdunik, J. L. and Haensel, P.},
  title = {Neutron drip transition in accreting and nonaccreting neutron star crusts},
  journal = {Phys. Rev. C},
  volume = {91},
  pages = {055803},
  year = {2015},
  doi = {10.1103/PhysRevC.91.055803}
}

@article{Hartle1967,
  author = {Hartle, James B.},
  title = {Slowly Rotating Relativistic Stars. {I}. Equations of Structure},
  journal = {Astrophys. J.},
  volume = {150},
  pages = {1005--1029},
  year = {1967},
  doi = {10.1086/149400}
}

@article{HartleThorne1968,
  author = {Hartle, James B. and Thorne, Kip S.},
  title = {Slowly Rotating Relativistic Stars. {II}. Models for Neutron Stars
           and Supermassive Stars},
  journal = {Astrophys. J.},
  volume = {153},
  pages = {807--834},
  year = {1968},
  doi = {10.1086/149707}
}

@article{ReinaVera2015,
  author = {Reina, Borja and Vera, Ra{\"u}l},
  title = {Revisiting {Hartle}'s model using perturbed matching theory to
           second order: amending the change in mass},
  journal = {Class. Quantum Grav.},
  volume = {32},
  number = {15},
  pages = {155008},
  year = {2015},
  doi = {10.1088/0264-9381/32/15/155008}
}

@article{Hinderer2008,
  author = {Hinderer, Tanja},
  title = {Tidal {Love} Numbers of Neutron Stars},
  journal = {Astrophys. J.},
  volume = {677},
  pages = {1216--1220},
  year = {2008},
  doi = {10.1086/533487}
}

@article{Hinderer2009Erratum,
  author = {Hinderer, Tanja},
  title = {Erratum: ``Tidal {Love} Numbers of Neutron Stars''
           (2008, {ApJ}, 677, 1216)},
  journal = {Astrophys. J.},
  volume = {697},
  pages = {964},
  year = {2009},
  doi = {10.1088/0004-637X/697/1/964}
}

@article{Schenk2002,
  author = {Schenk, A. K. and Arras, P. and Flanagan, {\'E}. {\'E}.
            and Teukolsky, S. A. and Wasserman, I.},
  title = {Nonlinear mode coupling in rotating stars and the
           {$r$}-mode instability in neutron stars},
  journal = {Phys. Rev. D},
  volume = {65},
  pages = {024001},
  year = {2002},
  doi = {10.1103/PhysRevD.65.024001}
}

@article{FullerLai2012,
  author = {Fuller, Jim and Lai, Dong},
  title = {Dynamical tides in compact white dwarf binaries:
           tidal synchronization and dissipation},
  journal = {Mon. Not. R. Astron. Soc.},
  volume = {421},
  number = {1},
  pages = {426--445},
  year = {2012},
  doi = {10.1111/j.1365-2966.2011.20320.x}
}

@article{FullerLai2014,
  author = {Fuller, Jim and Lai, Dong},
  title = {Dynamical tides in compact white dwarf binaries: influence of rotation},
  journal = {Mon. Not. R. Astron. Soc.},
  volume = {444},
  number = {4},
  pages = {3488--3500},
  year = {2014},
  doi = {10.1093/mnras/stu1698}
}

@article{YuWeinbergFuller2020,
  author = {Yu, Hang and Weinberg, Nevin N. and Fuller, Jim},
  title = {Non-linear dynamical tides in white dwarf binaries},
  journal = {Mon. Not. R. Astron. Soc.},
  volume = {496},
  number = {4},
  pages = {5482--5502},
  year = {2020},
  doi = {10.1093/mnras/staa1858}
}

@article{TakatsyKovacs2020,
  author = {Tak{\'a}tsy, J{\'a}nos and Kov{\'a}cs, P{\'e}ter},
  title = {Comment on ``Tidal {Love} numbers of neutron and self-bound quark stars''},
  journal = {Phys. Rev. D},
  volume = {102},
  pages = {028501},
  year = {2020},
  doi = {10.1103/PhysRevD.102.028501}
}

@article{Ryan1995,
  author = {Ryan, Fintan D.},
  title = {Gravitational waves from the inspiral of a compact object into a
           massive, axisymmetric body with arbitrary multipole moments},
  journal = {Phys. Rev. D},
  volume = {52},
  pages = {5707},
  year = {1995},
  doi = {10.1103/PhysRevD.52.5707}
}

@article{McNeillMardlingMueller2020,
  author = {McNeill, L. O. and Mardling, R. A. and M{\"u}ller, B.},
  title = {Gravitational waves from dynamical tides in white-dwarf binaries},
  journal = {Mon. Not. R. Astron. Soc.},
  volume = {491},
  number = {2},
  pages = {3000--3012},
  year = {2020},
  doi = {10.1093/mnras/stz3215}
}

@article{Otoniel2026Torsion,
  author = {Otoniel, Edson and Rebou{\c{c}}as, Jonathan A. and Lobo, Iarley P.},
  title = {Teleparallel torsion and white dwarf structure in {$f(T)=T+\Xi T^2$} gravity},
  journal = {Physics Letters B},
  year = {2026},
  doi = {10.1016/j.physletb.2026.141113},
  eprint = {2609.11921},
  archivePrefix = {arXiv},
  primaryClass = {astro-ph.SR}
}

@article{Astashenok2023,
  author = {Astashenok, A. V. and Odintsov, S. D. and Oikonomou, V. K.},
  title = {Chandrasekhar Mass Limit of White Dwarfs in Modified Gravity},
  journal = {Symmetry},
  volume = {15},
  number = {6},
  pages = {1141},
  year = {2023},
  doi = {10.3390/sym15061141},
  eprint = {2211.14892},
  archivePrefix = {arXiv}
}

@article{Astashenok2022,
  author = {Astashenok, A. V. and Odintsov, S. D. and Oikonomou, V. K.},
  title = {Maximal masses of white dwarfs for polytropes in {$R^2$} gravity and theoretical constraints},
  journal = {Phys. Rev. D},
  volume = {106},
  number = {12},
  pages = {124010},
  year = {2022},
  doi = {10.1103/PhysRevD.106.124010},
  eprint = {2209.13693},
  archivePrefix = {arXiv}
}

@article{NojiriOdintsov2006,
  author = {Nojiri, Shin'ichi and Odintsov, Sergei D.},
  title = {Introduction to Modified Gravity and Gravitational Alternative for Dark Energy},
  journal = {eConf},
  volume = {C0602061},
  pages = {06},
  year = {2006},
  doi = {10.1142/S0219887807001928},
  eprint = {hep-th/0601213},
  archivePrefix = {arXiv}
}

@article{Capozziello2006,
  author = {Capozziello, S. and Nojiri, S. and Odintsov, S. D. and Troisi, A.},
  title = {Cosmological viability of {$f(R)$}-gravity as an ideal fluid and its compatibility with a matter dominated phase},
  journal = {Phys. Lett. B},
  volume = {639},
  pages = {135--143},
  year = {2006},
  doi = {10.1016/j.physletb.2006.06.034},
  eprint = {astro-ph/0604431},
  archivePrefix = {arXiv}
}

\end{document}